\documentstyle[12pt]{article}

\advance\hoffset by -1.1truecm
\advance\voffset by -1.0truecm

\def\be{\begin{equation}}
\def\ee{\end{equation}}
\def\ba{\begin{eqnarray}}
\def\ea{\end{eqnarray}}
\def\v#1{\vert #1 \rangle}
\def\scp#1#2{\langle#1\vert#2\rangle}

\def\me#1#2#3{\langle #1 \vert #2 \vert #3 \rangle}

\newcommand{\R}{\mbox{I \hspace{-0.82em} R}}
\newcommand{\N}{\mbox{I \hspace{-0.82em} N}}
\newcommand{\x}{{\bf x}}
\newcommand{\p}{{\bf p}}
\newcommand{\fsp}{{\bf sp}}
\newcommand{\etab}{{\bar{\eta}}}
\newcommand{\detab}{{\partial_{\etab}}}
\newcommand{\deta}{{\partial_{\eta}}}
\newcommand{\xl}{{\frac{x}{L}}}
\newcommand{\sn}{\smallskip\newline}
\newcommand{\mn}{\medskip\newline}

\newcommand{\mbo}{{\mbox{ }}}
\newcommand{\qqq}{ \qquad \qquad \qquad }
\def\detapb#1{{\partial_{\etab^{#1}}}}
\def\detapp#1{{\partial_{\eta^{#1}}}} 
     
\begin{document}

\title{On Quantum Field Theory with Nonzero Minimal Uncertainties 
in Positions and Momenta}
\author{Achim Kempf\thanks{Research Fellow of Corpus Christi 
College in the University of Cambridge}\\ 
Department of Applied Mathematics \& Theoretical Physics\\
University of Cambridge, Cambridge CB3 9EW, U.K.\\
 {\small Email: a.kempf@damtp.cambridge.ac.uk}}

\date{}
\maketitle

\begin{abstract}
We continue studies on quantum field theories on 
noncommutative geometric spaces, focusing on classes 
of noncommutative geometries which imply ultraviolet and infrared 
modifications in the form of nonzero minimal uncertainties in 
positions and momenta. The case of the ultraviolet modified 
uncertainty relation which has appeared from string theory and 
quantum gravity is covered. The example of euclidean $\phi^4$-theory 
is studied in detail and in this example we can now show 
ultraviolet and infrared regularisation of all graphs.

\end{abstract}

\vskip-14truecm
\hskip11truecm
{\bf DAMTP/96-22}\newline
$ \mbox{ \quad }$\hskip11.1truecm hep-th/9602085
\vskip14truecm

\newpage

\section{Introduction}
There has been considerable progress in several branches of the
mathematics of noncommutative or `quantum' geometry which, in a
broad sense, is the generalisation of geometric concepts and tools 
to situations in which the algebra of functions on a manifold becomes
noncommutative. The physical motivations range e.g. from 
integrable models and generalised symmetry groups to studies on the 
algebraic structure of the Higgs sector in the standard model. 
Standard references are e.g. \cite{sweedler}-\cite{connes-book}.
\sn
Here, we continue the approach of \cite{ak-lmp-bf}-\cite{ct} 
in which is studied the quantum mechanics on certain `noncommutative 
geometries' where 
\be 
[\x_i,\x_j] \ne 0 \mbox{\quad and \quad} [\p_i,\p_j] \ne 0 
\label{ab}
\ee
and in particular where:
\be
[\x_i,\p_j] = i\hbar ( \delta_{ij} + \alpha_{ijkl} \x_k\x_l +
\beta_{ijkl} \p_k\p_l + ... )
\label{ba}
\ee
A crucial feature of the generalised commutation relations, 
which we will discuss in Sec.2, is that for 
appropriate matrices $\alpha,\beta \in M_{n^4}(\bf C \rm)$ one finds 
ordinary quantum mechanical behaviour at medium scales, while as a new 
effect at very small and very large scales there appear nonzero minimal 
uncertainties $\Delta x_0, \Delta p_0$ in positions and in momenta. 
\sn
The main part of the paper is Sec.3, where we 
proceed with the study of a previously suggested approach 
to the formulation of quantum 
field theories on such geometries. For the example of $\phi^4$-theory 
we can now explicitly show that minimal uncertainties 
in positions and momenta do have the power to regularise 
all graphs in the ultraviolet and the infrared.
\sn
The underlying motivation is 
the idea is that nonvanishing minimal uncertainties in positions
and momenta could be effects caused by gravity, or string
theory. The possible gravitational origins for 
modifications in the ultraviolet and in the infrared are to be considered 
separately:
\sn
On the one hand, in order to resolve small distances test 
particles need high energies. The latest at the Planck scale
of about $10^{-35}m$ the gravity effects of 
high energetic test particles
must significantly disturb the spacetime structure which 
was tried to be resolved. It has therefore long been suggested that 
there exists a finite limit to the possible resolution of distances. 
Probably the simplest ansatz for its quantum theoretical expression 
is that of a nonvanishing minimal uncertainty in positions. This
ansatz covers an ultraviolet behaviour which has been 
found in string theory, as well as in quantum gravity, arising 
from an effective uncertainty relation:
\be
\Delta x \ge \frac{\hbar}{\Delta p} + \mbox{const} \cdot \Delta p
\label{ucrlit}
\ee
References are e.g. \cite{townsend}-\cite{maggiore}; 
a recent review is \cite{garay}.
\sn
On the other hand, minimal uncertainties in momentum, as an infrared 
effect, may arise from large scale gravity. The argument is related 
to the fact that on a general curved spacetime
there is no notion of a plane wave, i.e. of exact localisation in 
momentum space, see \cite{ak-ft,ct}.
\sn
We remark that in the case of minimal uncertainties in positions only,
examples are known of noncommutative geometries of the type of 
Eqs.\ref{ab},\ref{ba} which preserve the Poincar{\'e} symmetry, 
i.e. where the universal enveloping algebra of the Poincar{\'e} Lie
algebra is a $*$- sub algebra of the Heisenberg algebra, see
\cite{ak-gm-rm-prd,ak-gm-rm-2},
Generally however, we take the view that 
similarly to curved spaces which may preserve some of the flat space 
symmetries while breaking others, also 
noncommutative geometric spaces, as defined through commutation
relations, may preserve some symmetries while breaking others.
\sn
Here, we therefore study the general case, 
i.e. not assuming a specific symmetry, and allowing the existence both of 
minimal uncertainties in positions and in momenta.
\sn
An alternative approach with a similar motivation, 
but based on the canonical formulation of quantum field theory, is 
\cite{doplicher1}. Other approaches to nonrelativistic 
quantum mechanics with generalised commutation relations, mostly 
motivated by quantum groups, and related studies, are e.g. 
\cite{sm-cqg-plsc}-\cite{lukierski}.

\section{Quantum mechanics with nonzero minimal uncertainties}

\subsection{Uncertainty relations}
We review and generalise the results of \cite{ak-lmp-bf}-\cite{ak-jmp-ucr} 
on nonrelativistic quantum mechanics with 
nonzero minimal uncertainties in positions and momenta. 

Let ${\cal{A}}$ denote the associative Heisenberg algebra generated by 
elements $\x_i,\p_j$ that obey generalised commutation relations of the
form of Eqs.\ref{ab},\ref{ba}. The modified commutation
relations are required to be consistent with the $*$ - involution $\x^* = \x, \p^* = \p$, 
implying that $\alpha$ and $\beta$ obey
$\alpha^*_{ijkl} = \alpha_{ijlk}, \beta^*_{ijkl} = \beta_{ijlk}$.
\sn
The study of the uncertainty relations that belong to the Heisenberg 
algebra ${\cal{A}}$ yields information 
that holds independently of the choice of representation.
Let us therefore assume the $\x_i,\p_j$ to be represented as symmetric 
operators obeying the new commutation relations on some dense domain 
$D \subset \cal{H}$ in a Hilbert space $\cal{H}$. On this space $D$
of physical states one derives uncertainty relations
of the form
\be
\Delta A \Delta B \ge 1/2 \vert\langle[A,B]\rangle\vert 
\ee
so that e.g. $[\x_i,\x_j]\ne 0$, yields $\Delta x_i \Delta x_j \ge 0$. 
Their noncommutativity implies that the $\x_i$ (as well as the 
$\p_i$) can no longer be simultaneously diagonalised.
Because of the modified commutation relations Eqs.\ref{ba} and
the corresponding uncertainty relations there can appear the even more
drastic effect that the $\x_i$ (as well as the $\p_j$) may also not be 
diagonalisable separately. Instead there then exist nonzero minimal 
uncertainties in positions and momenta. The mechanism can be seen 
also in one dimension, to which case we will restrict ourselves until 
Sec.\ref{ndim}. We consider 
Eq.\ref{ba} with $\alpha,\beta > 0$ and $\alpha \beta < 1/\hbar^2$:
\begin{equation}
[\x,\p] = i \hbar (1 + \alpha \x^2 + \beta \p^2)
\label{1dimcr}
\end{equation}
For fixed but sufficiently small
$\alpha$ and $\beta$ one finds ordinary quantum mechanical behaviour at 
medium scales while e.g. the term proportional to $\beta$ contributes 
for large $\langle \p^2 \rangle = \langle \p \rangle ^2 + (\Delta p)^2$  
i.e. in the ultraviolet. Similarly the term proportional to $\alpha$
leads to an infrared effect. The uncertainty relation to Eq.\ref{1dimcr}
is:
\begin{equation}
\Delta x \Delta p \ge \frac{\hbar}{2} \left( 1 + \alpha (\Delta x)^2 
+ \alpha \langle \x\rangle ^2
+ \beta (\Delta p)^2 + \beta \langle \p\rangle ^2 \right)
\label{1dimucr}
\end{equation}
It implies nonzero minimal uncertainties in $\x$- as well as in $\p$- 
measurements. This can be seen as follows: As e.g. $\Delta x$ gets 
smaller, $\Delta p$ must increase so that the product $\Delta x \Delta p$ 
of the LHS remains larger than the
RHS. In usual quantum mechanics this is always possible, i.e. 
$\Delta x$ can be made arbitrarily small. However, in the
generalised case, for $\alpha,\beta>0$ there is a positive 
$(\Delta p)^2$ term on the
RHS which eventually grows faster with $\Delta p$ than the LHS.
Thus $\Delta x$ can no longer become arbitrarily small. 
The minimal uncertainty in $\x$ depends 
on the expectation value in position and momentum via \be
k:= \alpha \langle \x \rangle^2 + \beta \langle \p \rangle^2
\label{k}
\ee
and is explicitly:
\begin{equation}
\Delta x_{0} = \sqrt{\frac{(1+k)\beta \hbar^2}{1-\alpha\beta\hbar^2}}
\label{nmux}
\ee
Analogously one obtains the smallest uncertainty in momentum
\be
\Delta p_{0} = \sqrt{\frac{(1+k)\alpha \hbar^2}{1-\alpha\beta\hbar^2}}
\label{nmup}
\end{equation}
with the absolutely smallest uncertainties obtained for $k=0$.
\sn
Note that if there was e.g. an $\x$- eigenstate $\v{\psi} \in D$ 
with $\x.\v{\psi}=\lambda\v{\psi}$ it would
have no uncertainty in position (we always assume states $\v{\psi}$
to be normalised):
\be
(\Delta x)^2_{\v{\psi}} =
\langle \psi \vert (\x - \langle \psi \vert 
\x \vert \psi \rangle)^2 \v{\psi } = 0
\label{itw}
\ee
which would be a contradiction. There are thus no physical states 
$\vert \psi \rangle \in D$ which are eigenstates of $\x$ or $\p$. 
\sn
Thus, for any physical domain $D$, i.e. for all $*$-representations of the
commutation relations, there are no physical 
states in the `minimal uncertainty gap': 
\be
\exists\!\!\!/ \quad \vert \psi \rangle \in D: \quad
0 \le (\Delta x)_{\vert \psi \rangle} < \Delta x_0
\ee
\be
\exists\!\!\!/ \quad \vert \psi \rangle \in D: \quad
0 \le (\Delta p)_{\vert \psi \rangle} < \Delta p_0
\ee
Crucially, unlike on ordinary geometry, there do not 
exist sequences $\{\vert \psi_n \rangle\}$ of 
physical states which would approximate point localisations in 
position or momentum space, i.e. for which
the uncertainty would decrease to zero:
\be
\exists\!\!\!/ \quad \v{\psi_n} \in D: \qquad 
\lim_{n\rightarrow \infty} (\Delta x)_{\vert \psi_n \rangle} = 0. 
\mbox{\qquad or \qquad} 
\lim_{n\rightarrow \infty} (\Delta p)_{\vert \psi_n \rangle} = 0. 
\ee
Heisenberg algebras ${\cal{A}}$ with these generalised canonical 
commutation relations therefore no longer have spectral 
representations on wave functions $\scp{x}{\psi}$ or $\scp{p}{\psi}$.

\subsection{Bargmann Fock representation}
For practical calculations and for detailed studies of the functional
analysis a Hilbert space representation of the
generalised Heisenberg algebra is needed. 
We generalise the Bargmann Fock representation.
\mn
In ordinary quantum mechanics the Bargmann Fock representation is 
unitarily equivalent to the position and the momentum representation,
being the spectral representation 
of the operator $\etab \detab \in {\cal{A}}$ where: 
\be
\etab := \frac{1}{2L} \x  - \frac{i}{2 K} \p \mbox{ \quad and \quad }
\detab := \frac{1}{2L} \x  + \frac{i}{2 K} \p
\label{etadef}
\ee
Here $L$ and $K$ are length and momentum scales, related by $LK=\hbar/2$. 
Thus $\etab$ and $\detab$ obey $\detab \etab -\etab \detab = 1$, which 
is of the form of a Leibniz rule and justifies the notation.  
One readily finds the countable set of eigenvectors 
$ \etab \detab \v{\etab^{n}} = n \v{\etab^{n}}$ 
with $ n=0,1,2,... $. With the definitions
$\v{a \etab^n + b \etab^m} := \v{a \etab^n} + \v{b \etab^m} $ and $
 a \v{\etab^n} := \v{a \etab^n}$
arbitrary states $\v{\psi}$ are written as polynomials 
or power series 
\be
\v{\psi} = 
\v{\sum_{r=0}^{\infty}\psi_r\frac{\etab^r}{\sqrt{r!}}} = \v{\psi(\etab)}
\ee
on which $\x$ and $\p$ are represented in terms of multiplication
and differentiation operators
\be 
\x = L(\etab + \detab)
\qquad \qquad
\p = i K (\etab - \detab)
\label{acxp}
\ee
The well known formula for the scalar product of states is 
\be
\scp{\psi}{\phi} = \frac{1}{2\pi i} \int d\eta d\etab\mbo
\overline{\psi(\etab)}\mbo  e^{-\etab \eta}\mbo \phi(\etab)
\label{spalt}
\ee
Here the $\psi(\etab)$ and $\phi(\etab)$ on the RHS are to be read
as polynomials or power series in ordinary complex variables rather
than as elements of ${\cal{A}}$.
\sn
A key observation for the
generalisation of the Bargmann Fock representation is that the
scalar product can be expressed without relying to complex integration
\cite{ak-lmp-bf}:
\be
\scp{\psi}{\phi} = \overline{\psi(\etab)}\mbox{ }
e^{\partial_{\eta} \partial_{\etab}}\mbox{ }  
\phi(\etab)\mbo \vert_{\eta=0=\etab}
\label{spneu}
\ee
The exponential is defined through its power series i.e.
$e^{\partial_{\eta} \partial_{\etab}} =
\sum_{r=0}^{\infty}
\frac{\partial_{\eta} \partial_{\etab}}{r!}$
where the derivatives $\detab$ act from the left while 
the derivatives $\deta$ act from the right.
The evaluation procedure is to carry out the differentiations
and then to set $\eta$ and $\etab$ equal to zero. The remaining
number is the value of the scalar product. 
This can be done purely algebraically by using the Leibniz 
rule $\partial_{\etab}\etab - \etab \partial_{\etab} = 1$ and
its complex conjugate $\eta \partial_{\eta} - \partial_{\eta} \eta =1$. 
For example
\begin{eqnarray*} 
\detab \etab^2 &=& \detab \etab \etab = (\etab \detab +1)\etab
= \etab \detab \etab +\etab\\ 
 &=& \etab (\etab \detab +1) +\etab
=\etab \etab \detab + \etab+\etab = 2 \etab
\end{eqnarray*}
and
$$
\eta^2 \deta = \eta (\deta \eta +1) = ... = 2 \eta
$$
Thus e.g.:
\begin{eqnarray*}
\scp{\etab^2}{2+3\etab^2} &=& \eta^2\mbo 
e^{\partial_{\eta} \partial_{\etab}}\mbox{ }
(2+3\etab^2) \mbo \vert_{\eta=0=\etab}\\
 &=& \eta^2 \sum_{r=0}^{\infty}
\frac{\partial_{\eta} \partial_{\etab}}{r!} 
\mbo (2+3\etab^2)
\mbo \vert_{\eta=0=\etab}\\
 &=& 3 \eta^2 \frac{\deta^2\detab^2}{2} 
\etab^2\mbo \vert_{\eta=0=\etab} = 6
\end{eqnarray*}
Since the scalar product formula Eq.\ref{spalt} relies on
conventional commutative integration over the complex plane, it cannot
be used in the generalised case where e.g. in $n$ dimensions
the $\etab_i$ will be noncommutative.
It is however possible to use a generalisation (Eq.\ref{qsp})
of Eq.\ref{spneu} (which can also be applied in the fermionic case instead of 
using Berezin integration \cite{ak-lmp-bf}). Also in one dimension
it allows to construct a Bargmann Fock Hilbert space 
representation for Eq.\ref{1dimcr}.
\sn
To this end we rewrite Eq.\ref{1dimcr} in the form
\begin{equation}
[\x,\p] = i \hbar + i \hbar (q^2 -1) 
\left( \frac{\x^2}{4 L^2}+\frac{\p^2}{4 K^2} \right)
\label{1dimcrqLK}
\end{equation}
where the parameter $q \ge 1$ measures the deviation from the ordinary 
commutation relations. The length and momentum scales
are related by $LK=\hbar(q^2+1)/4$. We can now again represent $\x$ and
$\p$ as the usual linear combinations (Eq.\ref{acxp}) of
generators $\etab$ and $\detab$. A complete 
generalised Bargmann Fock calculus is defined as the complex associative
algebra $\cal{B}$ with the commutation relations
\be
\detab \etab - q^2 \etab \detab = 1 \qquad \qquad
\eta \deta - q^2 \deta \eta = 1
\label{1dbfa}
\ee
\be
\etab \deta - q^2 \deta \etab = 0 \qquad \qquad 
\detab \eta - q^2 \eta \detab = 0
\label{1dbfee}
\ee
\be
\eta \etab -q^2 \etab \eta = 0 
\qquad \qquad \deta \detab -q^2 \detab \deta = 0
\label{1dbfe}
\ee
A short calculation shows that the commutation relation Eq.\ref{1dimcrqLK}
in fact uniquely translates into the commutation relations 
Eqs.\ref{1dbfa} through Eq.\ref{acxp}, see [13]. On the other hand,
the commutation relations Eqs.\ref{1dbfee},\ref{1dbfe} are nonunique
and could also be chosen commutative. Our choice is 
the special case of the choice made for the $n$ dimensional 
case in [10] under the requirements of a quantum group module 
algebra structure, invariance of the Poincar{\'e} series and 
simple form of the scalar product formula. 
These requirements are here not physically relevant, but it is 
convenient to use the formulas already obtained for this case.
Generally, other choices for the commutation relations 
between the barred and the unbarred generators are possible and 
lead to respectively more or less simple to evaluate formulations
of the scalar product. These representation specific choices do 
of course not affect the physical content of the theory, 
such as the uncertainty relations, transition amplitudes or 
expectation values.
\sn
The Heisenberg algebra ${\cal{A}}$ is now represented on the 
domain $D$ of polynomials in $\etab$ 
\be
D:= \left\{ \v{\psi} \vert \quad \psi(\etab) = \mbox{polynomial}(\etab) 
\right\}
\ee
with the action of $\x$ and $\p$ given by Eq.\ref{acxp}
where the differentiations are to be evaluated algebraically
using the generalised Leibniz rule given in Eqs.\ref{1dbfa}. 
As is the case on ordinary geometry, the operators $\etab$ and 
$\detab$ are mutually adjoint with respect 
to the unique and positive definite scalar product, which now takes the form:
\be
\scp{\psi}{\phi} = \overline{\psi(\etab)}\mbo
e_{1/q}^{\partial_{\eta}\partial_{\etab}}
\mbo \phi(\etab) \mbo \vert_{\eta=0=\etab}
\label{qsp}
\ee
The $q$- exponential is defined through
\be                                 
e_{1/q}^{\partial_{\eta}\partial_{\etab}}
= \sum_{r=0}^{\infty}\frac{( 
\partial_{\eta_i}\partial_{\etab_i})^r}{[r]_{1/q}!}
\ee
where the derivatives $\deta$ act from the right and where
$$
[r]_{c} := 1 + c^{2} + c^{4} + ... + c^{2(r-1)} = 
\frac{c^{2r}-1}{c^{2}-1}
$$
and 
$$
[r]_c! := 1\cdot [2]_c\cdot [3]_c\cdot ... \cdot [r]_c
$$
The evaluation procedure is again to algebraically 
carry out the differentiations, now using Eqs.\ref{1dbfa}-\ref{1dbfe}
and then to set $\eta$ and $\etab$ equal to zero. The remaining
number is the value of the scalar product. 

The functional analysis of the position and  momentum operators 
is as follows: 
We denote by $H$ the Hilbert space obtained by completion
with respect to the norm induced by the scalar product. 
A Hilbert basis is given by the orthonormal family
\be
\left\{ ([r]_q!)^{-1/2} \v{\etab^r} \quad \vert \quad r=0,1,2,...\right\}
\ee
The domain $D \subset H$, which is dense in $H$, is a physical domain, i.e. 
on it the $\x$ and $\p$ are represented as symmetric operators obeying 
the commutation relation Eq.\ref{1dimcrqLK}. In fact $D$ is also
analytic since $\x.D \subset D$ and $\p.D \subset D$, i.e.
$D$ is a $*$-${\cal{A}}$ module.
The $\x$ and $\p$ are no longer essentially self-adjoint.
Their adjoints $\x^*$ and $\p^*$ are closed but nonsymmetric.
The $\x^{**}$ and $\p^{**}$ are closed and symmetric.
Their deficiency subspaces are of finite (nonzero) and equal dimension
so that there are continuous families of self adjoint extensions in $H$. 
Crucially however, because of the minimal uncertainties
in positions and momenta, neither $\x$ nor $\p$ have self-adjoint extensions
neither in $D$ nor in any other physical domain, 
i.e. not in any other $*$-representation of 
the commutation relations. For the details and proofs see \cite{ak-jmp-ucr}. 
\sn 
One arrives at the following picture:
\sn
While in classical mechanics the states can 
have exact positions and momenta, in quantum mechanics there is the 
uncertainty relation that does not allow $\x$ and $\p$ to have common 
eigenvectors. Nevertheless
$\x$ and $\p$ separately do have 'eigenvectors', 
though non-normalisable ones. The
spectrum is continuous, namely the configuration or momentum
space. The position and momentum operators are essentially self-adjoint.
Our generalisation of the Heisenberg algebra
has further consequences for the observables $\x$ and $\p$: 
It is not only that the $\x$ and $\p$
have no common eigenstates. 
The uncertainty relation now implies that they do not have any
eigenvectors in the representation
of the Heisenberg algebra. Although $\x$ and $\p$ 
separately do have self-adjoint extensions, they do not have self-adjoint
extensions on any physical domain i.e. not on any $*$-representation
of both $\x$ and $\p$. This means the non-existence of absolute 
precision in position or momentum measurements. Instead there are 
absolutely minimal uncertainties in these measurements which are, in terms
of the new variables of Eq.\ref{1dimcrqLK}:
\be 
\Delta x_0 = L \sqrt{1-q^{-2}}, \qquad \qquad 
\Delta p_0 = K \sqrt{1-q^{-2}}
\label{nmun}
\ee
Recall that due to Eq.\ref{itw} the non self-adjointness 
and non-diagonalisability of $\x$ and $\p$ is  
necessary to allow for the physical description of minimal 
uncertainties. Note that on the other hand the fact that $\x$ and $\p$ 
still have the slightly weaker property of being symmetric 
is sufficient to guarantee that all physical expectation 
values are real. 

\subsection{Maximal localisation states}
Generally, all information on positions and momenta is 
encoded in the matrix elements of the position and momentum operators, and
matrix elements can of course be calculated in any basis. 
In the Bargmann Fock basis matrix elements e.g. of the position
operators are calculated as
\be
\me{\psi}{\x}{\phi} = \overline{\psi(\etab)}\mbo
e_{1/q}^{\deta\detab}\mbo L(\etab + \detab)\mbo \phi(\etab)
\mbo \vert_{0}
\ee
Ordinarily, information on position or momentum
can conveniently be obtained by projection onto position or momentum 
eigenstates $\scp{x}{\psi}$ or $\scp{p}{\psi}$ i.e. by using a position or 
momentum representation.
\sn
That there are now no more physical $\x$- or $\p$- eigenstates, 
can also be seen directly in the Bargmann Fock representation. 
We consider e.g. the eigenvalue problem for $\x$
\be
\x.\v{\psi_{\lambda}} = \lambda \v{\psi_{\lambda} }
\mbox{ \qquad i.e. \qquad } L(\etab + \detab) 
\psi_{\lambda}(\etab)
= \lambda \psi_{\lambda}(\etab)
\label{recev}
\ee
which yields a recursion formula for the coefficients of the expansion:
\be
\psi_{\lambda}(\etab) = \sum_{r=0}^{\infty} \psi_{\lambda,r} \etab^r
\ee
In ordinary quantum mechanics
the solution is a Dirac $\delta$ 'function', transformed into
Bargmann Fock space, (i.e. Eq.\ref{bfdelta} with $\lambda$ instead of
$x_0$). In the generalised setting it is interesting to see 
the effect of the appearance of the minimal uncertainty `gap'.

The (no longer generally mutually orthogonal) solutions $\sum_{r=0}^{\infty}
\psi_{\lambda,r} \etab^r$ to Eq.\ref{recev} have 
vanishing uncertainty in positions but they are not contained
in the domain of $\p$ (this would of course 
contradict the uncertainty relation) and they are 
therefore not physical states.
However every polynomial approximation to the power series 
is contained in the physical domain $D$, i.e.
$\sum_{r=0}^{n} \psi_{\lambda,r} \etab^r \in D$ for arbitrary finite $n$.
Thus, each $\sum_{r=0}^{n} \psi_{\lambda,r} \etab^r$ has 
an $\x$- uncertainty which is in fact larger than $\Delta x_0$.
For details and a graph of their scalar product see \cite{ak-jmp-ucr}.
\sn
Let us now consider the physical states 
$\v{\phi^{mlx}_{\xi,\pi}}$, $\v{\phi^{mlp}_{\xi,\pi}}$
which have the maximal localisation 
in $\x$ or $\p$ for given expectation values $\xi,\pi$ in positions
and momenta:
\be
\Delta x_{\v{\phi^{mlx}_{\xi,\pi}}} = \Delta x_0
\ee
\be
\langle \phi^{mlx}_{\xi,\pi} \vert \x \v{\phi^{mlx}_{\xi,\pi}} = \xi
\mbox{ , \qquad \quad }
\langle \phi^{mlx}_{\xi,\pi} \vert \p \v{\phi^{mlx}_{\xi,\pi} } = \pi
\ee
with $\Delta x_0$ given by Eq.\ref{nmux} 
and similarly for $\v{\phi^{mlp}_{\xi,\pi}}$.
E.g. the projection $\scp{\phi^{mlx}_{\xi,\pi} }{\psi}$ is then the
probability amplitude for finding the particle maximally localised
around $\xi$ with momentum expectation $\pi$. For 
$\alpha, \beta  \rightarrow 0$ one recovers the 
position and the momentum eigenvectors.
 
In order to calculate e.g. the $\v{\phi^{mlx}_{\xi,\pi}}$ we use that
these physical states realise the equality in the uncertainty relation.
As is well known the uncertainty relation 
follows from the positivity of the norm:
\be
\vert (\x -\langle \x \rangle + \frac{\langle [\x,\p]\rangle}{2(\Delta p)^2}
(\p-\langle \p \rangle))\v{\psi}\vert \ge 0
\ee
which is
\be
\langle \psi \vert (\x-\langle \x\rangle)^2 - \left(
\frac{\vert\langle[\x,\p]\rangle\vert}{2 (\Delta p)^2}\right)^2
(\p-\langle\p\rangle)^2 \v{\psi} \ge 0
\ee
so that:
\be
\Delta x \Delta p \ge \frac{\vert \langle[\x,\p]\rangle\vert}{2}
\ee
Thus, a state $\v{\psi}$ obeys
$\Delta x \Delta p = \vert \langle[\x,\p]\rangle\vert /2$,
i.e. it is on the boundary of the physically allowed region if:
\be
(\x -\langle \x \rangle + \frac{\langle [\x,\p]\rangle}{2(\Delta p)^2}
(\p-\langle \p \rangle))\v{\psi} = 0
\label{squeezed}
\ee         
In any given representation this equation has a family of squeezed state
solutions parametrized by 
$\langle \x\rangle, \langle \p\rangle, \Delta x, \Delta p$
where the four parameters obey Eq.\ref{1dimucr} with the equality sign.
Choosing for $\Delta x$ or $\Delta p$ the minimal values given by
 Eqs.\ref{nmux},\ref{nmup} yields the maximal localisation states
$\v{\phi^{mlx}_{\xi,\pi} } $ and $\v{\phi^{mlp}_{\xi,\pi} } $.
\sn
In \cite{ak-gm-rm-prd}, we calculated maximal localisation states in 
the case $\alpha =0$. The absence of a minimal uncertainty in 
momentum there allows a spectral representation of $\p$, with 
Eq.\ref{squeezed} taking the form of an exactly solvable differential 
equation. In particular, the new concept of quasi-position representation 
has been introduced, where the Heisenberg algebra is represented on the
wave functions $\psi(\xi) := \scp{\phi^{mlx}_{\xi,0}}{\psi}$. Related to the 
minimal uncertainty in positions there appears a minimal wavelength 
in quasi-position space.

In the general situation with minimal 
uncertainties in positions and in momenta
we work in Bargmann Fock space where Eq.\ref{squeezed} takes the form
(using Eq.\ref{k})
\be
\left(L(\etab+\detab)  
-\langle \x \rangle + i \hbar \frac{1+\alpha (\Delta x)^2
+ \beta (\Delta p)^2 + k}{2(\Delta p)^2}
(iK(\etab-\detab)-\langle \p \rangle) \right) \psi(\etab) = 0
\ee
yielding a three terms recurrence 
relation for the coefficients of the expansion
of $\v{\psi}$ in $\etab$. The solutions, 
i.e. the maximal localisation states, 
can be expressed in terms of so-called $q$-continuous Hermite functions.
A detailed study of the maximal localisation states and the corresponding 
quasi-position and quasi-momentum representation has been carried out 
in \cite{ak-hh-1}. A survey of $q$- special 
functions is \cite{koekoek-swarttouw}.

A further problem is to find a generalised Fourier transformation
that allows to easily transform information on positions 
into information on momenta. While this has been worked out for
the special case $\alpha =0$ in \cite{ak-gm-rm-prd}, here
the recent work \cite{ak-sm-jmp} may be relevant. 
In this context, compare also with the 
generalised quantum mechanics (with discrete $x$- and $p$- spectra)
developed in \cite{toy1,toy2}, where techniques developed in 
\cite{kw} lead to generalised Fourier transformations.

\subsection{Integral kernels and Green functions}
Elements $P = P(\x,\p) \in {\cal{A}}$ of the Heisenberg algebra 
do not only have representations in terms of Bargmann 
Fock operators $P(\etab,\detab)$, via Eq.\ref{acxp} but
can also still be represented as integral kernels. 
Once the operator $P(\etab,\detab)$ is normal ordered, 
there is a simple rule for deriving its integral kernel $G_P$, 
which is a function of 
${\bar{\eta}}^{\prime}$ and $\eta$. Integrating any Bargmann Fock function 
${\psi}(\bar{\eta})$ over $G_P({\bar{\eta}}^{\prime},{\eta})$ 
leads then to a 
function of ${\bar{\eta}}^{\prime}$, which is 
$P.{\psi}({\bar{\eta}}^{\prime})$. Generalising
\be
P(\etab^{\prime},\partial_{\etab^{\prime}}).\psi(\etab^{\prime})
= \frac{1}{2\pi i}
\int d\etab d\eta\mbo G_P(\etab^{\prime},\eta)\mbo
e^{-\etab \eta} \mbo \psi(\etab)
\ee
one now has
\begin{equation}
P(\etab^{\prime},\partial_{\etab^{\prime}}).\psi(\etab^{\prime}) =
\int d{\bar{\eta}} d{\eta}\mbo  G_P({{\bar{\eta}}}^{\prime},{\eta})\mbo
e_{1/q}^{{\partial}_{{\eta}}{\partial}_{\bar{\eta}}}  
\mbo {\psi}({\bar{\eta}})
\end{equation}
Here the integration is meant to be the algebraic scalar product which
expresses the integration in terms of derivatives, i.e. one defines:
\be
\int d{\bar{\eta}} d{\eta}\mbo  \overline{\psi(\etab)}\mbo
 e_{1/q}^{{\partial}_{{\eta}}{\partial}_{\bar{\eta}}}\mbo 
{\phi}({\bar{\eta}})
 := \overline{\psi(\etab)}
\mbo e_{1/q}^{{\partial}_{{\eta}}{\partial}_{\bar{\eta}}} 
\mbo \phi(\etab)\mbo \vert_{\eta=0=\etab}
\ee
For this to work, 
the appropriate commutation relations between two copies (e.g. primed and
unprimed) of the function
space had to be calculated, see \cite{ak-jmp-bf}.
\sn
E.g. the position operator $\x := L (\etab + \detab)$, has 
the integral kernel
\be
G_x({{\bar{\eta}}^{\prime}},\eta ) =
L\mbo (\etab^{\prime}\mbo 
e_{1/q}^{{{\bar{\eta}}^{\prime}}{\eta}} +
\mbo e_{1/q}^{{{\bar{\eta}}^{\prime}}{\eta}}\mbo \eta )
\ee
Another example is the harmonic oscillator $H := \omega \etab \detab$. 
Since $H$ is self-adjoint, the time evolution operator $U = e^{-i(t_f-t_i)H}$
is unitary. The eigenvalues of $H$ are:
\be
H\mbo \v{\etab^r} = \omega \mbo [r]_q \v{\etab^r}
\ee
The integral kernel of $U$, i.e. the Greens function is then
found to be \cite{ak-jmp-bf}:
\be
G_U =
\sum_{r=0}^{\infty} 
\frac{ { ( {{\bar{\eta}}^{\prime}}{\eta})}^r}{[r]_{1/q}!}
e^{-i\omega (t_f-t_i)[r]_q}
\label{eagf}
\ee
reducing for $q \rightarrow 1$ to the well known result:
\be
G_U(\etab^{\prime},\eta) = 
e^{\etab^{\prime}\eta e^{-i\omega (t_f-t_i)}}
\ee
\subsection{$n$- dimensional  generalisations}  
\label{ndim}
Let us come back to the full $n$- dimensional situation with
commutation relations of the form of Eqs.\ref{ab},\ref{ba}. 
Obviously, terms $\alpha_{ijii} > 0$ 
and $\beta_{ijii}>0$ are sufficient to
induce minimal uncertainties in momenta and
positions, thus excluding spectral representations 
of the $\x_i$ or $\p_j$, and therefore complicating the 
construction of Hilbert space representations of the Heisenbarg algebra.
\sn
There are however $n$- dimensional generalisations of our $q$- Bargmann 
Fock space which straightforwardly supply Hilbert space representations for 
certain classes of generalised Heisenberg algebras. 
We will use two of them as examples of fixed `background' geometries.

The first example is the Heisenberg algebra ${{\cal{A}}}_1$, defined as 
the tensor product of $n$ commuting copies of the one-dimensional 
algebra $\cal{A}$ (all $q_i\ge 1$):
\begin{equation}
[\x_i,\p_j] = i \hbar \delta_{ij} + i \hbar \delta_{ij} (q_i^2-1)
 \left(\frac{1}{4L_i^2} \x_i^2 + \frac{1}{4 K_i^2} \p_i^2
\right)
\label{2Hi}
\end{equation}
\be
[\x_i,\x_j] = 0
, \qqq
[\p_i,\p_j] = 0
\ee
where 
\be
L_i K_i = \hbar (q_i^2 +1)/4
\label{2Hf}
\ee
The Heisenberg algebra ${\cal{A}}_1$ has an obvious Hilbert space 
representation on the domain $D_1 \subset H_1$ which is the $n$ fold tensor
product of the previously considered domains $D$ 
in the Hilbert space $H_1$, spanned by the orthogonal polynomials
$\etab_1^{r_1} \etab_2^{r_2}\cdot ... \cdot \etab_n^{r_n}$, 
with norm:
\be
\scp{\etab_1^{r_1} \etab_2^{r_2}\cdot ... \cdot \etab_n^{r_n}}{
\etab_1^{r_1} \etab_2^{r_2}\cdot ... \cdot \etab_n^{r_n}}
= \prod_{i=1}^{n} [r_i]_{q_i}!
\ee
\sn
As the second example, now with nontrivial 
commutation relations also among the $\x_i$ and
among the $\p_i$ we consider the Heisenberg algebra ${\cal{A}}_2$
defined through:
\begin{equation}
[\x_r,\p_r] = i \hbar + i \hbar (q^2 -1) 
\sum_{s\le r} 
\left( \frac{q^2+1}{2}\right)^{s-1}
\left(\frac{\x_s^2}{4 L_s^2}+\frac{\p_s^2}{4 K_s^2} \right)
\label{xpcr}
\label{1Hi}
\end{equation}
and mixed commutation relations for $s>r$
\begin{equation}
[\x_s,\p_r] =  -i \frac{K_r}{L_r} \frac{q-1}{q+1} \{ \x_s, \x_r\}  
\qqq 
[\x_s,\x_r] =  -i \frac{L_r}{K_r} \frac{q-1}{q+1} \{ \x_s, \p_r\}  
\end{equation}
and for $s<r$
\begin{equation}
[\x_s,\p_r]  =  i \frac{L_s}{K_s} \frac{q-1}{q+1} \{ \p_s, \p_r\}  
\qqq 
[\p_s,\p_r]  =  -i \frac{K_s}{L_s} \frac{q-1}{q+1} \{ \x_s, \p_r\}  
\label{1Hf}
\end{equation}
with:
\begin{equation}
L_r K_r := \frac{\hbar}{2} \left(\frac{q^2 +1}{2}\right)^r
\label{klpbez}
\end{equation}
In order to represent ${\cal{A}}_2$ we define the generalised 
Bargmann Fock calculus as the complex algebra ${\cal{B}}_2$ with 
commutation relations (the $i,j$ summed over):
\be
\etab_a \etab_b - \frac{1}{q} R_{ba}^{ji} \etab_j \etab_i = 0
\ee
\be
\detapb{a} \etab_b - q R_{ib}^{aj} \etab_j \detapb{i} = \delta_{ab}
\qqq
\detapb{a}\detapb{b} - \frac{1}{q} R_{ab}^{ij} \detapb{j}\detapb{i} =0
\label{twoa}
\ee
\be
\detapb{a}\detapp{b} - \frac{1}{q} (R^{-1})_{bi}^{ja} 
\detapp{j}\detab{i}=0
\qqq
\detapb{a}\eta_b - q R_{ab}^{ij} \eta_j \detapb{i} = 0
\ee
and their complex conjugates\footnote{Note that $\bar{ }$ is an
anti algebra morphism, so that e.g. 
$\overline{\detapb{i}\etab_j} = \eta_j \detapp{i}$ (we defined
the $\deta$'s as right derivatives)} where (the $e_i^j$ are matrix units):
\begin{equation}
R = q \sum_i e^i_i \otimes e^i_i + \sum_{i\ne j} e^i_i \otimes e^j_j + 
(q-1/q) \sum_{i>j} e^i_j \otimes e^j_i
\label{1bff}
\label{surm}
\ee
We can then represent ${\cal{A}}_2$ through
\begin{equation}
\x_r = L_r (\etab_r + \detapb{r}) \mbox{ \qquad and \qquad } 
\p_r = i K_r (\etab_r - \detapb{r})
\label{xpansatz} 
\end{equation}
on the domain of polynomials
\be
D_2:= \left\{ \v{\psi} \vert \quad \psi(\etab) = 
\mbox{polynomial}(\etab_1,\etab_2,...,\etab_n) \right\}
\ee
with the unique and positive definite scalar product  
\be
\scp{\psi}{\phi} = \overline{\psi(\etab)}\mbo
e_{1/q}^{\partial_{\eta_i}\partial_{\etab_i}}
\mbo \phi(\etab) \mbo \vert_{\eta=0=\etab}
\label{spgen}
\ee
which is a generalisation of Eq.\ref{spneu}. 
The Hilbert space $H_2$, completed with respect to the induced norm,
has a Hilbert basis given by the orthogonal ordered polynomials
$\etab_1^{r_1} \etab_2^{r_2}\cdot ... \cdot \etab_n^{r_n}$ with norm:
\be
\scp{\etab_1^{r_1} \etab_2^{r_2}\cdot ... \cdot \etab_n^{r_n}}{
\etab_1^{r_1} \etab_2^{r_2}\cdot ... \cdot \etab_n^{r_n}}
= \prod_{i=1}^{n} [r_i]_q!
\label{norma2}
\ee
The Heisenberg algebra ${\cal{A}}_2$ and its Hilbert space representation 
has naturally appeared in the context of quantum groups 
\cite{ak-lmp-bf}-\cite{ak-jmp-ucr} as a minimal generalisation under 
certain consistency conditions such as the invariance of the $*$- structure,
Poincar{\'e} series, and the positivity of the norm. 
$R$ in Eq.\ref{surm} is the fundamental representation of the 
universal R-matrix that determines the quasitriangular structure 
of the quantum group $SU_q(n)$, which 
acts on ${\cal{A}}_2$ as linear quantum canonical
transformations, i.e. ${\cal{A}}_2$ is a $SU_q(n)$- $*$- comudule algebra.

Generally, a Hilbert space representation of fixed generalised 
commutation relations induces Hilbert space representations
of a class of generalised commutation relations, simply
by applying algebra isomorphisms ($M\in GL(n,\bf R\rm)$):
\be
\x_r \rightarrow \x_r^{\prime} = M^{-1}_{rs} \x_s \qquad
\quad \p_r \rightarrow \p_r^{\prime} = M_{sr} \p_s
\label{aliso}
\ee
E.g. the noncommutative geometries ${\cal{A}}_1$,${\cal{A}}_2$ 
defined through Eqs.\ref{2Hi}-\ref{2Hf} and Eqs.\ref{1Hi}-\ref{1Hf} 
are of the form (summing over repeated indices):
\be
[\x_r,\p_s]  =  i\hbar \delta_{rs} + i\hbar \alpha_{rstu} \{ \x_t,\x_u\}
+ i\hbar \beta_{rstu} \{\p_t,\p_t\} 
\label{genHi}
\ee
\be
[ \x_r ,\x_s]  =  i \mu_{rstu} \{ \x_t, \p_u \} 
\ee
\be
 [\p_r,\p_s]  =  i \nu_{rs,tu} \{ \x_t, \p_u \}
\label{genHf}
\ee
with the $\alpha,\beta,\mu,\nu$ real matrices. 
Through Eqs.\ref{aliso} one represents commutation 
relations of the same form Eqs.\ref{genHi}-\ref{genHf} but specified through 
matrices $\alpha^{\prime},\beta^{\prime},\mu^{\prime},\nu^{\prime}$, where
\begin{eqnarray}
\alpha^{\prime}_{abcd} & = & M^{-1}_{ai} M_{jb} M_{kc} M_{ld} \alpha_{ijkl} 
\\
\beta^{\prime}_{abcd} & = & M^{-1}_{ai} M_{bj} M^{-1}_{ck} 
M^{-1}_{dl} \beta_{ijkl} 
\\
\mu^{\prime}_{abcd} & = & M^{-1}_{ai} M^{-1}_{bj} M_{kc} 
M^{-1}_{dl} \mu_{ijkl}
\\
\nu^{\prime}_{abcd} & = & M_{ia} M_{jb} M^{-1}_{ck} M_{ld} \nu_{ijkl}
\end{eqnarray}
Note that since unitary transformations generally preserve the 
commutation relations, the transformations Eqs.\ref{aliso} are 
noncanonical and lead to commutation relations that describe 
different physical behaviour. 

The two Heisenberg algebras ${\cal{A}}_1$ and ${\cal{A}}_2$ 
will also serve as examples for fixed background noncommutative geometries
in our quantum field theoretical studies.

\section{Quantum field theory with minimal uncertainties}
In Sec.3.1 a general approach to the path integral formulation 
of quantum field theories on noncommutative geometric spacetimes is applied.
As an example, euclidean $\phi^4$- theory is formulated in Sec.3.2
on the spacetimes ${\cal{A}}_1$ and ${\cal{A}}_2$, using the 
previously developed Bargmann Fock space techniques.
The structure constants of the pointwise multiplication 
of fields are calculated in Sec.3.3. The Feynman rules are 
derived in Sec.3.4 and, using their asymptotic behaviour it
is shown in Sec.3.5 that, on the spacetimes ${\cal{A}}_1$ and ${\cal{A}}_2$,
all graphs of $\phi^4$-theory are regularised.

\subsection{Path integral on noncommutative geometric spaces}
In the euclidean path integral formulation 
a field theory is defined through its partition function.
\be
Z = N \int_F D\phi\mbo e^{-1/\hbar \mbo S[\phi]}
\ee
where $N$ is a normalisation constant and
\be
S: F \rightarrow \R \qquad 
S: \phi \rightarrow S[\phi]
\ee
is a nonlinear action functional from the space $F$ of fields 
to the real numbers. 
\sn
The space $F$ of fields $F \subset H$ in a Hilbert space $H$
is a $*$-representation of the Heisenberg algebra ${\cal{A}}$ 
generated by elements $\x_{i}$ and $\p_{j}$, ordinarily obeying
\be
[\x_{i},\p_{j}] = i\hbar \delta_{i,j} 
\label{ordcr}
\ee
The closure of $F$ under addition insures the translation 
invariance of the path integral. The $\p_i$ act on fields 
e.g. in the kinetic action, while the $\x_i$ act on 
fields e.g. in gauge transformations $\psi \rightarrow 
\exp(i \alpha(\x)).\psi$. 
\sn
Of course, 
in quantum field theory the generators $\x_i$ and $\p_j$ of the Heisenberg 
algebra $\cal{A}$ do no longer have 
the simple quantum mechanical interpretation as 
observables of positions and momentum, because of 
the existence of antiparticles.
Nevertheless, positions and momenta do not become mere parameters 
in quantum field theory. It is this Heisenberg algebra $\cal{A}$ which 
is setting the quantum theoretical stage of position and momentum spaces, 
also in quantum field theory, see 
also e.g. \cite{feynman-qed,feynman in dirac lecture}.
\sn
Generally, the action functionals $S$ of local field theories 
can be expressed in terms of the action of $\cal{A}$ on fields $\phi \in F$,
where $F$ is a $*$-representation, the scalar product 
$\fsp(\mbo,\mbo)$ in $F$, 
and the pointwise multiplication `$*$' of fields:
\be
* \mbo : \quad F \otimes F \rightarrow F
\ee
Let us consider the example of charged $\phi^4$ theory:
\be
Z[J] := N \int D\phi\mbo e^{\int d^4x\mbo
\phi^* (\partial_i\partial_i - \mu^2)\phi 
 - \frac{\lambda}{4!}(\phi \phi)^*\phi \phi + \phi^*J+J^*\phi}
\ee
Here velocities and actions are measured as multiples of $c$ and 
$\hbar$. Reintroducing the fundamental constants,
together with a unit length $l$, yields:
\be
Z[J] = N \int D\phi\mbo D\phi^*\mbo e^{\int d^4x\mbo 
\frac{-l^2}{\hbar^2} \phi^* ( \p_i\p_i + m^2 c^2).\phi 
 - \frac{\lambda l^4}{4!}(\phi\phi)^*\phi\phi + \phi^*J+J^*\phi}
\label{pi2}
\ee
The choice of $l$ does 
not affect the theory since it can be absorbed in a finite redefinition 
of the fields and the coupling constant. It will of course drop out 
of the Feynman rules.
\sn
The Heisenberg algebra $\cal{A}$ defined by Eq.\ref{ordcr} acts on the
fields as
\be
\x_i.\phi(x) = x_i \phi(x) \qqq \p_j.\phi(x) 
= -i\hbar \partial/\partial_{x_j} \phi(x),
\ee
we define a scalar product $\fsp(\mbo,\mbo)$ in $F$ 
\be
\fsp(\phi_1,\phi_2) = \int d^4x \mbo \phi_1^*(x) \phi_2(x)
\ee
and the pointwise multiplication $*:\mbo F \otimes F \rightarrow F$:
\be 
(\phi_1 * \phi_2)(x) = \phi_1(x) \phi_2(x)
\ee
Since we require the space $F$ of fields $\phi$ 
that is to be summed over in the
path integral to be a $*$- representation of the commutation 
relations Eq.\ref{ordcr}, 
a suitable specification of $F$ is $F:= S_{\infty} \subset H:=L^2$. 
The domain $F$ is an analytic ($F$ is a $*$-$\cal{A}$ module)
and dense domain in the Hilbert space $H$ of square integrable functions.  
\mn
The pointwise multiplication $*$ equipes $F$ with the structure
of a non-unital commutative algebra. $F$ is closed under the 
associative multiplication,
while the identity $\phi(x) \equiv 1$ is neither in $F$ nor in $H$.
The commutativity of $*$
\be
\forall \mbo \phi_1,\phi_2 \in F: \qquad\phi_1 * \phi_2 = \phi_2 *\phi_1
\label{ax1}
\ee
is crucial for the description of 
bosons and can (and will) be preserved on the noncommutative geometries.
\sn
The above definitions yield:
\be
Z[J] = N \int_F D \phi \mbo e^{ - \frac{l^2}{\hbar^2} \fsp\left( 
\phi, 
\mbo ( \p^2 + m^2 c^2  ).\phi \right) \mbo
- \frac{\lambda l^4}{4!} 
\fsp\left( \phi * \phi, \phi * \phi \right) \mbo
+ \fsp(\phi,J) + \fsp(J,\phi) }
\label{pi1}
\ee
The units are now fully transparent since, through the 
introduction of $l$, the abstract 
fields $\phi \in F$ do not carry units.
Their pointwise product $\phi_1 * \phi_2$ does carry units.
\sn
Eq.\ref{pi1} provides a formulation of the path integral which 
is independent of the choice of a Hilbert basis in $F \subset H$. From 
Eq.\ref{pi1} one obtains Eq.\ref{pi2} 
by choosing the spectral representation of the position operators $\x_i$.
Equivalently one may choose other Hilbert bases in $H$, such as
e.g. the spectral representation of the momenta in which 
the Heisenberg algebra $\cal{A}$ acts on the fields 
as $\x_i.\phi(p) = i \hbar \partial/\partial_{p_i} \phi(p)$ and
$\p_j.\phi(p) = p_j \phi(p)$ 
with the scalar product $\fsp(\mbo,\mbo)$ in $F$ reading
\be
\fsp(\phi_1,\phi_2) = \int d^4p \mbo \phi_1^*(p) \phi_2(p)
\ee
and the pointwise multiplication $*:\mbo F \otimes F \rightarrow F$ 
taking the form of the convolution product:
\be 
(\phi_1 * \phi_2)(p) = 
(2\pi\hbar)^{-2} \int d^4k\mbo \phi_1(k) \phi_2(p-k)
\label{convol}
\ee
The form of Eq.\ref{pi1} is not only representation independent.
It is crucial that it does also not rely on fixed 
commutation relations in the Heisenberg algebra $\cal{A}$. 
\mn
Our approach to the formulation of quantum field theories on noncommutative 
geometries is therefore to stick to the
abstract form of the action functional, as e.g. in Eq.\ref{pi1},
while generalising the Heisenberg algebra $\cal{A}$.
This means a generalisation of the `stage' of space-time and energy-momentum 
on which the field theory is built, technically through 
changes in the action of the operators on fields, 
the scalar product and in the pointwise product of fields, which are then 
reflected in the Feynman rules. Note that the scalar products could
be written as traces, using $\fsp(a,b) = \sum_n\fsp(n,b)\mbo\fsp(a,n)=
\mbox{\bf tr}( \vert b)(a\vert)$, with $\{\vert n)\}_n$ being a Hilbert 
basis in $H$.

\subsection{$\phi^4$-theory on the geometries ${\cal{A}}_1$ 
and ${\cal{A}}_2$}
The framework can be applied for the formulation of quantum field theories on
generic noncommutative background geometries which may or
may not have certain symmetries, similar to the case of curved background
geometries. Here, we will use the nontrivial examples of non Lorentz 
symmetric noncommutative background geometries ${\cal{A}}_1$ and 
${\cal{A}}_2$ (e.g. for $n=4$), since they are known to imply minimal 
uncertainties and since we can conveniently make use of our previous 
results on the construction of explicit Hilbert space representations.

Note that the quantum mechanics for Lorentz symmetric examples 
of suitable noncommutative background geometries 
was studied in \cite{ak-gm-rm-prd} and the
corresponding field theoretical studies are in progress \cite{ak-gm-rm-2}.
\mn
Generally, for practical calculations a representation is needed on a 
domain $F$ of fields in a Hilbert space $H$ and a Hilbert basis to work in.
Neither ${\cal{A}}_1$ nor ${\cal{A}}_2$
have spectral representations of the $\x_i$ or $\p_j$, while the
Bargmann Fock representations on the domains $F := D_1$ or $D_2$,
as developed in Sec.2, can again be used.
Fields are given as polynomials or power series $\phi(\etab_1,...
\etab_n)$ rather than as functions $\phi(x)$ or $\phi(p)$,
with the action of the operators $\x_i,\p_j$ given by Eq.\ref{xpansatz}.
\sn
The abstract action functional of Eq.\ref{pi1} is to be 
expressed, term by term, in the Bargmann Fock representation.
\mn
In the case of ${\cal{A}}_2$ the scalar product of fields reads, from 
Eq.\ref{spgen}:
\be
\fsp(\phi_1,\phi_2) = \overline{\phi_1(\etab)}\mbo
e_{1/q}^{\partial_{\eta_i}\partial_{\etab_i}}
\mbo \phi_2(\etab) \mbo \vert_0
\ee
Here and in the following we sum over repeated indices and 
$\vert_0$ stands for 'all differentiations evaluated at zero'.
\sn
The source terms are scalar products:
\be
\fsp(\phi,J) = \overline{\phi(\etab)}\mbo
e_{1/q}^{\deta_i\detab_i}\mbo 
J(\etab)\vert_0 
\ee
\be
\fsp(J,\phi) = \overline{J(\etab )} \mbo 
e_{1/q}^{\deta_i\detab_i}\mbo 
\phi(\etab)\vert_0 
\ee From Eq.\ref{pi1} the free part of the action functional is the 
scalar product of the field $\phi$ with the field $Q.\phi$:
\be
S_{0}[\phi] = \fsp(\phi, Q.\phi)
\ee
where
\be
Q := \frac{l^2}{\hbar^2}( \p_i\p_i+m^2c^2) 
\label{skin}
\label{Qa}
\ee
which acts on Bargmann Fock space as:
\be 
\frac{l^2}{\hbar^2}(\p_i\p_i +m^2c^2).\phi(\etab) 
= \frac{l^2}{\hbar^2}
\left(-\sum_{i=1}^{4}K_i^2 (\etab_i-\detab_i)^2 + m^2c^2 \right)\phi(\etab)
\label{Qb}
\ee
Thus, the free action reads:
\begin{eqnarray}
S_{0}[\phi] & = & \frac{l^2}{\hbar^2} \mbo
\fsp\left(\phi, \left( \p_i\p_i 
+ m^2c^2 \right).\phi\right) 
\nonumber \\ 
  & = & \frac{l^2}{\hbar^2} \mbo
  \overline{\phi(\etab )}\mbo e_{1/q}^{\deta_i\detab_i}
\left(-\sum_{i=1}^{4}K_i^2 (\etab_i-\detab_i)^2 
+ m^2 c^2 \right)\phi(\etab) \vert_0
\end{eqnarray}
The interaction term is the scalar product 
of the field $\phi * \phi$ with itself, it thus reads
in Bargmann Fock space:
\begin{eqnarray}
S_{int}[\phi] & = & \frac{\lambda l^4}{4!} 
\mbo \fsp(\phi * \phi, \phi * \phi) 
\nonumber \\
  & = & \frac{\lambda l^4}{4!} \mbo \overline{(\phi*\phi)(\etab )}
\mbo e_{1/q}^{\deta_i\detab_i}\mbo
(\phi *\phi)(\etab) \vert_0
\label{vertex}
\end{eqnarray}
These are the expressions for ${\cal{A}}_2$. In the case of the geometry 
${\cal{A}}_1$ the exponential is replaced by the product of exponentials 
\be
e_{1/q}^{\sum_{i=1}^{4} \deta_i\detab_i} \mbo \rightarrow \mbo
\prod_{i=1}^{4} e_{1/q_i}^{\partial_{\eta_i}\partial_{\etab_i}}
\label{e1e2}
\ee
while the case of ordinary geometry is of course recovered 
for $q$ or all $q_i \rightarrow 0$. 
\mn
Recall that the $\etab_i$ have two multiplicative structures, related to
to the Heisenberg algebra ${\cal{A}}$ and to the algebra $F$ of fields.
Solving Eqs.\ref{xpansatz} for $\etab_i$,
the $\etab_i$ act as multiplication operators
on the fields and can be identified with elements of the Heisenberg
algebra, thus, in the generalised case, reflecting its noncommutativity.
On the other hand the fields $\phi(\etab)$ are commutatively 
multiplied pointwise, through `$*$' for the description of local interaction. 
\sn
The structure constants $C_{\vec{r},\vec{s},\vec{t}}$
(here and in the following index `vectors' $\vec{r}$ take 
values $\vec{r} \in \N^4$)
\be
C_{\vec{r},\vec{s},\vec{t}} := \fsp(\etab_1^{r_1}\cdot ... \cdot
\etab_4^{r_4},\etab_1^{s_1}\cdot ... \cdot \etab_4^{s_4} 
* \etab_1^{t_1}\cdot ... \cdot \etab_4^{t_4})
\label{stco}
\ee
will be needed explicitly.

\subsection{Pointwise multiplication}
On ordinary geometry the pointwise multiplication $*$ 
transforms into momentum space as the well known convolution product:
\begin{eqnarray}
(\phi * \phi)(x) & =& \phi(x)\phi(x)\\
(\phi * \phi)(p) & =& (2\pi\hbar)^{-1/2} \int_{-\infty}^{+\infty}
dk\mbo \phi_p(k) \phi_p(p-k)
\label{conv}
\end{eqnarray}
In order to obtain the convolution product formula, two arbitrary 
functions on momentum space are unitarily (Fourier-) 
transformed into position space, 
multiplied pointwise, and the resulting function is unitarily
(Fourier-) transformed back into momentum space, yielding Eq.\ref{conv}.
\sn
Analogously the unitary equivalence of the 
Bargmann Fock- with the position space
representation determines the pointwise multiplication 
in Bargmann Fock space and the $C_{\vec{r},\vec{s},\vec{t}}$ uniquely:
\sn
The matrix elements of the unitary transformation to the 
spectral representation of $\x$ are
\be
\scp{x}{\etab^n} = \sqrt{n!} (2\pi L^2)^{-1/4} (x/2L - L\partial_x)^n 
e^{-\frac{1}{4}\left(\frac{x}{L}\right)^2}
\label{unha1}
\ee
i.e., up to a factor, the Hermite functions. 
The use of Eq.\ref{unha1} for the
transformation of $*$ from position space into Bargmann Fock space 
is however rather inconvenient. Starting from known 
expressions, more practical formulas can be developed.
\sn
As has been known since \cite{bargmann} (see also \cite{perelomov} and
references therein), fields $\phi(x)$ given in the position representation 
are transformed into the Bargmann Fock representation by 
\be
\phi(\etab) = (2\pi L^2)^{-1/4} \int_{-\infty}^{+\infty} dx\mbox{ }
e^{-\frac{1}{2}\etab^2+\etab\xl -\frac{1}{4}(\xl)^2} \phi(x)
\label{g1}
\ee
with the inverse:
\be
\phi(x) = (8\pi^3L^2)^{-1/4}\int_{-i\infty}^{+i\infty} d\etab \mbox{  }
e^{\frac{1}{2}\etab^2-\etab\xl +\frac{1}{4}(\xl)^2} \phi(\etab)
\label{g2}
\ee
To see this, note that the Bargmann Fock function $\phi(\etab):=1$ is mapped
onto
\be
\phi(x)=(2\pi L^2)^{-1/4}\mbo e^{-1/4\mbox{ }(x/L)^2}
\ee
and vice versa. 
The induction is then completed by showing that multiplying the
Bargmann Fock function with $\etab$ amounts to the action
of $(x/L -2L\partial_x)/2$ on the field in position space.
\sn
These formulas, connecting the position space with the Bargmann Fock
space, are analogues of the Fourier transformation formulas connecting
the position space with the momentum space. 
Similar formulas connect Bargmann Fock space directly
to momentum space:
\be
\phi(\etab) = \left(\frac{2L^2}{\pi \hbar^2}\right)^{1/4} 
\int_{-\infty}^{+\infty} dp\mbo e^{\frac{1}{2}\etab^2 +
2i\etab\frac{Lp}{\hbar}-\left(\frac{Lp}{\hbar}\right)^2} \phi_p(p)
\label{mombf}
\ee
with the inverse:
\be
\phi_p(p) = \left(\frac{L^2}{2\pi^3 \hbar^2}\right)^{1/4} 
\int_{-\infty}^{+\infty} d\etab\mbo e^{-\frac{1}{2}\etab^2 -
2i\etab\frac{Lp}{\hbar}+\left(\frac{Lp}{\hbar}\right)^2} \phi(\etab)
\label{bfmom}
\ee
For the proof, note that $\fsp(p,\phi(\etab)=1) = 
\left(\frac{2L^2}{\pi \hbar^2}\right)^{1/4} 
e^{-\left(\frac{Lp}{\hbar}\right)^2}$.
\sn
Let us remark that from Eqs.\ref{g1},\ref{g2} immediately follows
that the transformation
\be
\tilde{f}(y):=\int_{-\infty}^{+\infty} dx \mbox{ } 
e^{-\frac{1}{L^2}(x-y)^2} f(x)
\label{uns}
\ee
which yields a `Gau{\ss}ian-diluted' function has an inverse:
\be
f(x)=\frac{1}{\pi L^2} \int_{-\infty}^{+\infty} dy\mbox{ }
e^{\frac{1}{L^2}(x-iy)^2} \tilde{f}(iy)
\label{revuns}
\ee
The $\x$-eigenvector with eigenvalue $x_0$, i.e. in position space 
the `$\delta$- function' at $x_0$, has the Bargmann Fock representation
$\phi_{(x_0)}(\etab)$ (using Eq.\ref{g1}):
\be
\phi_{(x_0)}(\etab) =
(2\pi L^2)^{-1/4} e^{-\frac{\etab^2}{2} + 
\etab \frac{x_0}{L} - \frac{1}{4}(\frac{x_0}{L})^2}
\label{bfdelta}
\ee
The scalar product of an arbitrary $\phi(\etab)$ with $\phi_{(x_0)}(\etab)$ 
yields another formula for the transformation from 
Bargmann Fock to position space, using Eq.\ref{spalt}:
\be
\phi(x) = \frac{(2\pi L^2)^{-1/4}}{2\pi i} \int d\eta d\etab 
e^{-\etab \eta -\frac{\eta^2}{2}+\eta \xl 
-\frac{1}{4}(\xl)^2} \phi(\etab)
\label{rueck2}
\ee
Similarly, the use the algebraic form Eq.\ref{spneu} of the scalar
product yields
\be
\phi(x) = (2\pi L^2)^{-1/4} \mbo 
e^{-\frac{\eta^2}{2} + \eta\xl
-\frac{1}{4}(\xl)^2}\mbox{ }
e^{\partial_{\eta}\partial_{\etab}}\mbox{ } 
\phi(\etab) \vert_{\eta=0=\etab}
\ee
and thus:
\be
\phi(x) = (2\pi L^2)^{-1/4} \mbo 
e^{-\frac{1}{2}\partial_{\etab}^2 + \xl \partial_{\etab}
-\frac{1}{4}(\xl)^2}\mbox{ }
\phi(\etab) \vert_{\etab=0}
\label{algebraischeruecktrafo}
\ee
This new transformation formula no longer involves integrations and can 
be evaluated algebraically, using the Leibniz rule only. 
\mn
The pointwise multiplication on Bargmann Fock space is now calculated by
unitarily transforming two arbitrary Bargmann Fock functions 
into position space, using the new formula Eq.\ref{algebraischeruecktrafo},
multiplying pointwise, and unitarily transforming the resulting function 
back into Bargmann Fock space, using Eq.\ref{g1}, to obtain:
\begin{eqnarray}
(\phi_1 * \phi_2)(\etab) & = & 
(2\pi L^2)^{-\frac{3}{4}} \int_{-\infty}^{+\infty} dx\mbo
e^{-\frac{1}{2} (\etab^2 + \partial^2_{\etab^{\prime}} +
\partial^2_{\etab^{\prime \prime}}) +
\frac{x}{L} (\etab + \partial_{\etab^{\prime}} +
\partial_{\etab^{\prime \prime}})
- \frac{3}{4} (x/L)^2 }
\phi_1(\etab^{\prime}) \phi_2(\etab^{\prime \prime}) \vert_0
\nonumber \\
 &   & \nonumber  \\
 & = & \left(\frac{2}{9\pi L^2}\right)^{\frac{1}{4}}
e^{\frac{1}{3}(\etab + \partial_{\etab^{\prime}} +
\partial_{\etab^{\prime \prime}})^2 -
\frac{1}{2} (\etab^2 + \partial^2_{\etab^{\prime}} +
\partial^2_{\etab^{\prime \prime}})}
\mbo \phi_1(\etab^{\prime}) \phi_2(\etab^{\prime \prime}) \vert_0
\label{star1dim}
\end{eqnarray}
This is the convolution product formula for Bargmann Fock space.
It allows to calculate the $C_{rst}$:
\begin{eqnarray}
C_{rst} & = & \fsp(\etab^r,\etab^s * \etab^t) \nonumber \\
 & = & \eta^r e^{\deta \detab} 
\mbo  \left(\frac{2}{9\pi L^2}\right)^{\frac{1}{4}}
e^{\frac{1}{3}(\etab + \partial_{\etab^{\prime}} +
\partial_{\etab^{\prime \prime}})^2 -
\frac{1}{2} (\etab^2 + \partial^2_{\etab^{\prime}} +
\partial^2_{\etab^{\prime \prime}})}
\mbo \etab^{\prime}{}^s \etab^{\prime \prime}{}^t \vert_0
\nonumber \\ 
  & = & \left(\frac{2}{9\pi L^2}\right)^{\frac{1}{4}}
\mbo e^{\frac{1}{3}(\detab + \partial_{\etab^{\prime}} +
\partial_{\etab^{\prime \prime}})^2 -
\frac{1}{2} (\detab^2 + \partial^2_{\etab^{\prime}} +
\partial^2_{\etab^{\prime \prime}})}
\mbo \etab^r \etab^{\prime}{}^s \etab^{\prime \prime}{}^t \vert_0
\end{eqnarray}
Using 
\be
\partial_x^r e^{a x + b x^2} \vert_{x=0} = \sum_{s \le r/2}
\frac{r!}{s! (r-2s)!} a^{r-2 s} b^s
\ee
we evaluate
\begin{eqnarray}
 &   & 
e^{\frac{-1}{6}(\detab^2 + \partial_{\etab^{\prime}}^2 +
\partial_{\etab^{\prime \prime}})^2 + \frac{2}{3} 
(\detab \partial_{\etab^{\prime}} +
\detab \partial_{\etab^{\prime \prime}} 
+ \partial_{\etab^{\prime}} \partial_{\etab^{\prime \prime}} )} 
\etab^r \etab^{\prime}{}^s \etab^{\prime \prime}{}^t \vert_0\nonumber \\
 & = & 
e^{\frac{-1}{6}( \partial_{\etab^{\prime}}^2 +
\partial^2_{\etab^{\prime \prime}}) + \frac{2}{3} 
(\partial_{\etab^{\prime}} \partial_{\etab^{\prime \prime}} )} 
\detab^r 
e^{\frac{-1}{6} \etab^2 + \frac{2}{3} 
\etab (\partial_{\etab^{\prime}} + \partial_{\etab^{\prime \prime}})} 
\etab^{\prime}{}^s \etab^{\prime \prime}{}^t \vert_0\nonumber \\
 & = & 
e^{\frac{-1}{6}( \partial_{\etab^{\prime}}^2 +
\partial^2_{\etab^{\prime \prime}}) + \frac{2}{3} 
(\partial_{\etab^{\prime}} \partial_{\etab^{\prime \prime}} )} 
\sum_u \frac{r!}{u!(r-2 u)!}\left( \frac{2}{3}(
\partial_{\etab^{\prime}} + \partial_{\etab^{\prime \prime}} 
)\right)^{r-2 u} \left(\frac{-1}{6}\right)^u
\etab^{\prime}{}^s \etab^{\prime \prime}{}^t \vert_0 \nonumber\\
 & = & 
e^{\frac{-1}{6}( \partial_{\etab^{\prime}}^2 +
\partial^2_{\etab^{\prime \prime}}) + \frac{2}{3} 
(\partial_{\etab^{\prime}} \partial_{\etab^{\prime \prime}} )} 
\sum_u \frac{r!\mbo \left( \frac{2}{3}\right)^{r-2u} 
\left(\frac{-1}{6}\right)^u
}{u!(r-2 u)!}
\sum_{v=0}^{r-2 u} \left(\matrix{r-2 u \cr v}\right)
\partial_{\etab^{\prime}}^{r-2 u -v}
\partial_{\etab^{\prime\prime}}^v \mbo
\etab^{\prime}{}^s \etab^{\prime \prime}{}^t \vert_0\nonumber \\
 & = & 
e^{\frac{-1}{6}( \partial_{\etab^{\prime}}^2 +
\partial^2_{\etab^{\prime \prime}}) + \frac{2}{3} 
(\partial_{\etab^{\prime}} \partial_{\etab^{\prime \prime}} )} 
\sum_{u,v} \frac{r!
\left( \frac{2}{3}\right)^{r-2u}
\left( \frac{-1}{6}\right)^u }{u! (r-2 u -v)! v!} 
\partial_{\etab^{\prime}}^{r-2 u -v}
\partial_{\etab^{\prime\prime}}^v \mbo
\etab^{\prime}{}^s \etab^{\prime \prime}{}^t \vert_0\nonumber \\
 & = & 
e^{\frac{-1}{6}( \partial_{\etab^{\prime}}^2 +
\partial^2_{\etab^{\prime \prime}}) + \frac{2}{3} 
(\partial_{\etab^{\prime}} \partial_{\etab^{\prime \prime}} )} 
\sum_{u,v} \frac{r! s! t!
\left( \frac{2}{3}\right)^{r-2u}
\left( \frac{-1}{6}\right)^u \mbo 
\etab^{\prime}{}^{s-r+2 u+v} \etab^{\prime \prime}{}^{t-v} }{u! 
(r-2 u -v)! v! (s-r+ 2u +v)! (t-v)!}
\vert_0\nonumber\\
 &   &  \nonumber \\
 &   & \mbox{which is, substituting $u$ by $a:=r-2 u-v$} \nonumber\\
 &   &  \nonumber \\
 & = & e^{\frac{-1}{6}( \partial_{\etab^{\prime}}^2 +
\partial_{\etab^{\prime \prime}}^2) + \frac{2}{3} 
(\partial_{\etab^{\prime}} \partial_{\etab^{\prime \prime}} )} 
\sum_{a,v} \frac{r! s! t!
\left( \frac{2}{3}\right)^{a+v}
\left( \frac{-1}{6}\right)^{\frac{r-v-a}{2}} }{
(\frac{r-v-a}{2})! v! a! (s-a)!(t-v)!}
\etab^{\prime}{}^{s-a} \etab^{\prime 
\prime}{}^{t-v} \vert_0\nonumber\\
 & = & \sum_{a,v} \frac{r! s! t!
\left( \frac{2}{3}\right)^{a+v}
\left( \frac{-1}{6}\right)^{\frac{r-v-a}{2}} }{
(\frac{r-v-a}{2})! v! a! (s-a)!(t-v)!}
\partial_{\etab^{\prime}}^{s-a}
e^{\frac{-1}{6}( \etab^{\prime}{}^2 +
\partial_{\etab^{\prime \prime}}^2) + \frac{2}{3} 
({\etab^{\prime}} \partial_{\etab^{\prime \prime}} )} 
\etab^{\prime \prime}{}^{t-v} \vert_0\nonumber\\
 & = & \sum_{a,v,w} \frac{r! s! t!
\left( \frac{2}{3}\right)^{a+v}
\left( \frac{-1}{6}\right)^{\frac{r-v-a}{2}} }{
(\frac{r-v-a}{2})! v! a! (s-a)!(t-v)!}
\frac{(s-a)!
\left( \frac{2}{3}\right)^{s-a-2 w}
\left( \frac{-1}{6}\right)^w}{(s-a-2 w)! w!}
\partial_{\etab^{\prime \prime}}^{s-a- 2 w}
e^{\frac{-1}{6} \partial_{\etab^{\prime \prime}}^2}
\etab^{\prime \prime}{}^{t-v} \vert_0\nonumber\\
 &   &  \nonumber \\
 &   & \mbox{and, replacing $w$ by $z:=s-a-2 w$} \nonumber\\
 &   &  \nonumber \\
 & = & \sum_{a,v,z} \frac{r! s! t!
\left( \frac{2}{3}\right)^{a+v+z}
\left( \frac{-1}{6}\right)^{\frac{r-v+s-2 a-z}{2}} }{
a! v! z! (\frac{r-a-v}{2})!
(\frac{s-a-z}{2})! (t-v-z)!}
e^{\frac{-1}{6} \partial_{\etab^{\prime \prime}}^2}
\etab^{\prime \prime}{}^{t-v-z} \vert_0\nonumber\\
 & = & \sum_{a,v,z} \frac{r! s! t!
\left( \frac{2}{3}\right)^{a+v+z}
\left( \frac{-1}{6}\right)^{\frac{r-v+s-2 a-z}{2}} }{
a! v! z! (\frac{r-a-v}{2})!
(\frac{s-a-z}{2})! (t-v-z)!}
\left( \frac{-1}{6}\right)^{\frac{t-v-z}{2}} \frac{(t-v-z)!}{
(\frac{t-v-z}{2})!}
\end{eqnarray}
to obtain eventually:
\be
C_{rst} = \left(\frac{2}{9\pi L^2}\right)^{\frac{1}{4}}
\sum_{i_1,i_2,i_3} \frac{r!\mbo s!\mbo t! \mbo
(-4)^{i_1+i_2+i_3} \mbo
(-6)^{-\frac{r+s+t}{2}} }{
i_1!\mbo i_2!\mbo i_3! \mbo (\frac{r-i_2-i_3}{2})!\mbo
(\frac{s-i_1-i_3}{2})!\mbo (\frac{t-i_1-i_2}{2})!}
\label{Crst}
\ee
In the sum over the $i_1,i_2,i_3$ only those terms contribute for which 
the arguments of all factorials are positive integers, which is a finite
number of terms. 
\sn
Let us also consider an alternative pointwise multiplication $*^{\prime}$, 
which is infrared modified:
\be
(\phi *^{\prime} \phi)(x) := \phi(x) \phi(x)\mbo
e^{\frac{1}{4} \left(\frac{x}{L}\right)^2}
\ee
In Bargmann Fock space this now takes a simple form 
without the square of derivatives in the exponential:
\begin{eqnarray}
(\phi *^{\prime} \phi)(\etab) &=&
(2\pi L^2)^{-\frac{3}{4}} \int_{-\infty}^{+\infty} dx\mbo
e^{-\frac{1}{2} (\etab^2 + \partial^2_{\etab^{\prime}} +
\partial^2_{\etab^{\prime \prime}}) +
\frac{x}{L} (\etab + \partial_{\etab^{\prime}} +
\partial_{\etab^{\prime \prime}})
- \frac{1}{2} (x/L)^2 }
\phi(\etab^{\prime}) \phi(\etab^{\prime \prime}) \vert_0
\nonumber \\
 &=& (2\pi L^2)^{-\frac{1}{4}} 
e^{\etab \partial_{\etab^{\prime \prime}} + (\etab + 
\partial_{\etab^{\prime \prime}}) \partial_{\etab^{\prime}}}
\mbo \phi(\etab^{\prime}) \phi(\etab^{\prime \prime}) \vert_0
\nonumber \\
&=& (2\pi L^2)^{-\frac{1}{4}}
\phi(\etab + \partial_{\etab^{\prime}})\phi(\etab + \etab^{\prime})
\vert_0
\label{starp1dim}
\end{eqnarray}
Thus
\begin{eqnarray}
C_{rst}^{\prime} & = & \fsp(\etab^r, \etab^s * \etab^t) \nonumber \\
  & = & \eta^r \mbo e^{\deta \detab} \mbo (2\pi L^2)^{-\frac{1}{4}} 
e^{\etab \partial_{\etab^{\prime \prime}} + (\etab + 
\partial_{\etab^{\prime \prime}}) \partial_{\etab^{\prime}}}
\mbo \etab^{\prime}{}^s \etab^{\prime \prime}{}^t \vert_0\nonumber\\
 &=& (2\pi L^2)^{-\frac{1}{4}} 
e^{\partial_{\etab^{\prime}} \partial_{\etab^{\prime \prime}} +  
( \partial_{\etab^{\prime}} + \partial_{\etab^{\prime \prime}}) \detab }
\mbo \etab^r \etab^{\prime}{}^s \etab^{\prime \prime}{}^t \vert_0
\nonumber \\
  & = & (2\pi L^2)^{-\frac{1}{4}}
  e^{\partial_{\etab^{\prime}} \partial_{\etab^{\prime \prime}}}
\mbo (\partial_{\etab^{\prime}} + \partial_{\etab^{\prime \prime}})^r
\mbo \etab^{\prime}{}^s \etab^{\prime \prime}{}^t \vert_0\nonumber \\
 & = & (2\pi L^2)^{-\frac{1}{4}}
 e^{\partial_{\etab^{\prime}} \partial_{\etab^{\prime \prime}}}
\mbo \sum_{a=0}^{r} \left( \matrix{r \cr a}\right) \mbo
\partial_{\etab^{\prime}}^a 
\partial_{\etab^{\prime \prime}}^{r-a} \mbo \etab^{\prime}{}^s 
\etab^{\prime \prime}{}^t \vert_0 \nonumber \\
  & = & (2\pi L^2)^{-\frac{1}{4}} \mbo
e^{\partial_{\etab^{\prime}} \partial_{\etab^{\prime \prime}}}
\mbo \left( \matrix{r \cr a}\right) 
\frac{s! \mbo t!}{(s-a)! (t-r+a)!} \mbo \etab^{\prime}{}^{(s-a)}
\etab^{\prime \prime}{}^{(t-r+a)} \vert_0 
\nonumber \\
& = & (2\pi L^2)^{-\frac{1}{4}} \mbo
\mbo \frac{r! \mbo s! \mbo t!}{\left(\frac{-r+s+t}{2}\right)!
\left(\frac{r-s+t}{2}\right)!\left(\frac{r+s-t}{2}\right)!}
\end{eqnarray}
whenever the arguments of all factorials are positive integers, and 
zero otherwise.
Compare with the $C_{rst}$ which can be put into the form:
\be
C_{rst} = \left(\frac{2}{9\pi L^2}\right)^{\frac{1}{4}}
\sum_{m_i} \frac{r! s! t! \mbo (-4)^{m_1+m_2+m_3}
(-6)^{-\frac{r+s+t}{2}} }{
(\frac{m_1+m_2-m_3}{2})! (\frac{m_1-m_2+m_3}{2})! (\frac{-m_1+m_2+m_3}{2})! 
(\frac{r-m_1}{2})!
(\frac{s-m_2}{2})! (\frac{t-m_3}{2})!}
\ee
Recall that, in the position representation, the Bargmann Fock 
polynomials $\etab^m$ have the asymtotic behaviour (Eq.\ref{unha1}):
$\propto \exp(-(x/2L)^2)$.
The pointwise multiplication $*$ of the Hermite functions thus yields
a function of asymptotic behaviour 
$\propto \exp(-2 (x/2L)^2)$. The weighted multiplication
$*^{\prime}$ cancels one of the Gau{\ss}ian factors and thus keeps
the asymptotic behaviour unchanged under the multiplication.
Thus $F$ is closed also under $*^{\prime}$. While it has a modified 
(and divergent) infrared behaviour it is also 
strictly local and is commutative. 

\subsection{Feynman rules}
Generally, given a *-representation $F$ of a possibly generalised 
Heisenberg algebra $\cal{A}$, together with the structure
constants $C$ of the pointwise multiplication in this representation,
it is possible to evaluate the action functional for arbitray fields,
and to calculate the Feynman rules.
\sn
On the example background geometry ${\cal{A}}_2$
the fields and sources $\phi,J \in F$ are expanded
in the Hilbert basis given by the ordered orthonormal polynomials
\be
\phi(\etab)=\sum_{s_1,s_2,s_3,s_4=0}^{\infty} 
\phi_{s_1s_2s_3s_4}\mbo 
\frac{\etab_1^{s_1}\etab_2^{s_2}\etab_3^{s_3}\etab_4^{s_4}}{
\sqrt{[s_1]_q![s_2]_q![s_3]_q![s_4]_q!}}
\ee
\be
J(\etab)=\sum_{s_1,s_2,s_3,s_4=0}^{\infty} 
J_{s_1s_2s_3s_4}\mbo 
\frac{\etab_1^{s_1}\etab_2^{s_2}\etab_3^{s_3}\etab_4^{s_4}}{
\sqrt{[s_1]_q![s_2]_q![s_3]_q![s_4]_q!}}
\ee
so that fields $\phi \in F$ are represented by their coefficient vector
$\phi_{\vec{r}} := \phi_{r_1,r_2,r_3,r_4}$ with 
indices $r_i = 0,1,2,...\infty$, ($i=1,...,4$). 
\sn
In the case ${\cal{A}}_2$ the algebra generated by the $\etab_i$ is 
noncommutative and the nontrivial fact that the ordered polynomials 
still form a Hilbert basis is a consequence of the
invariance of the Poincar{\'e} series (i.e. of the dimensionalities
of the subspaces of polynomials of equal grade), which was one of the key 
conditions in the derivation of the generalised Bargmann Fock calculus,
see \cite{ak-lmp-bf}.
\mn
The coefficient matrix of the quadratic operator $Q$ (from 
Eqs.\ref{Qa},\ref{Qb}) in the free action functional
is obtained as
\be 
M_{\vec{r}\vec{s}} := \frac{l^2}{\hbar^2}
\frac{\overline{\etab_1^{r_1} \etab_2^{r_2} \etab_3^{r_3}
\etab_4^{r_4}}}{\sqrt{[r_1]_q! \cdot ... \cdot [r_4]_q!}}
\mbo e_{1/q}^{\deta_i\detab_i}\mbo
\left( m^2c^2-\sum_{i=1}^{4} K_i^2 (\etab_i - \partial_{\etab_i})^2
\right)
\frac{\etab_1^{s_1} \etab_2^{s_2} \etab_3^{s_3}
\etab_4^{s_4}}{\sqrt{[s_1]_q! \cdot ... \cdot [s_4]_q!}}\mbo \vert_0
\label{M}
\ee
while the matrix elements of the interaction term 
read (from Eq.\ref{vertex})
\be
V_{\vec{t}\vec{u}\vec{v}\vec{w}} =
\frac{\overline{\etab_1^{t_1} \etab_2^{t_2} \etab_3^{t_3} \etab_4^{t_4}
* \etab_1^{u_1} \etab_2^{u_2} \etab_3^{u_3} \etab_4^{u_4}}
}{\sqrt{[t_1]_q! \cdot ... \cdot [t_4]_q! [u_1]_q! \cdot ... \cdot [u_4]_q!}}
\mbo e_{1/q}^{\deta_i\detab_i} \mbo  
\frac{{\etab_1^{v_1} \etab_2^{v_2} \etab_3^{v_3} \etab_4^{v_4}
* \etab_1^{w_1} \etab_2^{w_2} \etab_3^{w_3} \etab_4^{w_4}}
}{\sqrt{[v_1]_q! \cdot ... \cdot [v_4]_q! [w_1]_q! \cdot ... \cdot 
[w_4]_q!}}\mbo \vert_0
\ee
and, using Eq.\ref{stco}:
\be
V_{\vec{t}\vec{u}\vec{v}\vec{w}} = \sum_{z_1,...,z_4 = 0}^{\infty}
\frac{C_{\vec{t},\vec{u},\vec{z}} 
\mbo C_{\vec{z},\vec{v},\vec{w}} }{\prod_{i=1}^{4} [z_i]_q!
\sqrt{ [t_i]_q! 
[u_i]_q! [v_i]_q! [w_i]_q! }}
\label{Vtuvw}
\ee
The formulas given apply to the case ${\cal{A}}_2$. 
For ${\cal{A}}_1$, together with Eq.\ref{e1e2},
the $q$'s carry indices $q_1,...,q_4$.
\sn
Note that the path integration can be written as the
product of a \it countably \rm infinite number of integrations:
\begin{eqnarray}
N \int D\phi(x)\mbo D\phi^*(x) \mbo e^{-S[\phi(x),\phi^*(x)]} 
&=& N \int D\phi \mbo D\overline{\phi} \mbo 
e^{-S[\phi(\etab),\overline{\phi(\etab)}]}\\
 &=& N \int \prod_{r_1,r_2,r_3,r_4=0}^{\infty} d\phi_{r_1,r_2,r_3,r_4}
\mbo d\phi^*_{r_1,r_2,r_3,r_4} 
\mbo e^{-S[\phi_{\vec{r}},\phi^*_{\vec{r}}]}
\nonumber
\end{eqnarray}
This discretisation of the infinite number of 
ordinary integrations which form the path integral is not related 
to the issue of e.g. ultraviolet regularisation. 
On ordinary geometry it is merely a result of our choice 
of representation, which is unitarily equivalent to the
conventional representations of the Heisenberg algebra $\cal{A}$. 
Generally, we are simply making use of the fact that 
the Hilbert space $H$ is separable, i.e. that
$H$ has discrete bases.
\sn
The Feynman rules can be derived in the standard way, using
the generating functional Eq.\ref{pi1}, which now reads:
\be
Z[J] = N \int D\phi \mbo D\phi^* \mbo e^{-\mbo
\phi_{\vec{r}}^*\mbo M_{\vec{r}\vec{s}}
\mbo \phi_{\vec{s}}\mbo 
-\frac{\lambda l^4}{4!} 
V_{\vec{t}\vec{u}\vec{v}\vec{w}}
\phi_{\vec{t}}^* \phi_{\vec{u}}^* \phi_{\vec{v}} \phi_{\vec{w}} \mbo
+ \mbo \phi_{\vec{r}}^* J_{\vec{r}} \mbo+\mbo
J^*_{\vec{r}} \phi_{\vec{r}} }
\ee
Recall that each index vector denotes four indices, corresponding to the four
euclidean dimensions, e.g. 
$\vec{r} = (r_1,r_2,r_3,r_4)$ where each index is summed over, e.g.
$ r_2 = 0,1,2,...\infty$. Pulling the interaction term in front of the
integral yields:
\be
Z[J] = N e^{-\frac{\lambda l^4}{4!}
V_{\vec{r}\vec{s}\vec{t}\vec{u}}
\frac{\partial}{\partial J_{\vec{r}}}
\frac{\partial}{\partial J_{\vec{s}}} 
\frac{\partial}{\partial J^*_{\vec{t}}} 
\frac{\partial}{\partial J^*_{\vec{u}}}} 
\int D\phi\mbo D\phi^*
\mbo e^{-\mbo \phi_{\vec{r}}^*\mbo M_{\vec{r}\vec{s}}
\mbo \phi_{\vec{s}}\mbo +\mbo \phi_{\vec{r}}^* J_{\vec{r}} \mbo+\mbo
J^*_{\vec{r}} \phi_{\vec{r}}} 
\ee
In the discrete representation the functional 
derivatives become ordinary partial derivatives. 
Rearranging the remaining integrand
\be
Z[J] = N e^{-\frac{\lambda l^4}{4!}
V_{\vec{r}\vec{s}\vec{t}\vec{u}}
\frac{\partial}{\partial J_{\vec{r}}}
\frac{\partial}{\partial J_{\vec{s}}} 
\frac{\partial}{\partial J^*_{\vec{t}}} 
\frac{\partial}{\partial J^*_{\vec{u}}}} 
\int D\phi D\phi^*
 \mbo e^{-\mbo (\phi_{\vec{r}}^* -  J^*_{\vec{s}}
M^{-1}_{\vec{s}\vec{r}}) M_{\vec{r}\vec{t}}
(\phi_{\vec{t}} - M^{-1}_{\vec{t}\vec{u}} J_{\vec{u}})\mbo
+\mbo J^*_{\vec{r}} M^{-1}_{\vec{r} \vec{s}} J_{\vec{s}}}
\ee
the path integral can be absorbed in the overall constant:
\be
Z[J] = N^{\prime} e^{-\frac{\lambda l^4}{4!} 
V_{\vec{r}\vec{s}\vec{t}\vec{u}}
\frac{\partial}{\partial J_{\vec{r}}}
\frac{\partial}{\partial J_{\vec{s}}} 
\frac{\partial}{\partial J^*_{\vec{t}}} 
\frac{\partial}{\partial J^*_{\vec{u}}}} 
\mbo e^{J^*_{\vec{r}} M^{-1}_{\vec{r} \vec{s}} J_{\vec{s}}}
\ee
The calculation of graphs now involves loop summations
rather than loop integrations, with the Feynman rule
for the free propagator
\be
\Delta_0(\vec{a},\vec{b}) = M^{-1}_{\vec{a}\vec{b}}
\label{frfp}
\ee
and the lowest order vertex:
\be
\Gamma_0(\vec{a},\vec{b},\vec{c},\vec{d}) =
-\frac{\lambda l^4}{4!} V_{\vec{a}\vec{b}\vec{c}\vec{d}}
\label{Gamma0}
\ee
In graphs, each internal propagator $\Delta_0$ is
attached to two legs of a vertex $\Gamma_0$. While
the propagator carries a factor of $l^{-2}$, each leg of the vertex
carries a factor $l$. Thus, as it should be, the 
length scale $l$ drops out of the calculation. 
\sn
The Feynman rules could e.g. be applied to the calculation of the first order 
correction to the propagator, i.e. to the tadpole graph which now reads:
\be
\Delta(\vec{a},\vec{b}) = M^{-1}_{\vec{a}\vec{b}}
\mbo - \mbo \frac{\lambda L^4}{3!}  \mbo 
\sum_{\vec{r},\vec{s},\vec{t},\vec{u}}
V_{\vec{r}\vec{s}\vec{t}\vec{u}}\mbo
M^{-1}_{\vec{u}\vec{a}}
M^{-1}_{\vec{t}\vec{r}}
M^{-1}_{\vec{b}\vec{s}} \mbo + ...
\label{tad}
\ee
where e.g. $\sum_{\vec{r}}$ denotes 
$\sum_{r_1,r_2,r_3,r_4 = 0}^{\infty}$.
On ordinary geometry the tadpole contribution is divergent, since
it reads in momentum space, up to the external legs and a constant,
\be
\int d^4p\mbo \frac{1}{p_i p_i + m^2c^2} = 
\mbox{quadr. UV divergent}
\label{tadpolemom}
\ee
On ordinary geometry, i.e. with 
the ordinary Heisenberg algebra $\cal{A}$ underlying,
the Feynman rules in the Bargmann Fock representation 
are of course equivalent to those in the then existing 
position and momentum representations,
the change of Hilbert basis in $F$ is unitary, its determinant 
is trivial and no anomalies are introduced.
While $n$-point functions $\Gamma^{(n)}(\vec{x_1},...,\vec{x_n})$
and $\Gamma^{(n)}(\vec{p_1},...,\vec{p_n})$ are related by 
unitary (Fourier-) transformations they can also 
be transformed unitarily to and from the Bargmann Fock representation 
$\Gamma^{(n)}(\vec{r_1},...,\vec{r_n})$, 
using the transformations given in 
Eqs.\ref{g1},\ref{g2},\ref{mombf},\ref{bfmom}. On geometries
with minimal uncertainties there is still the possibility of
unitarily transforming to quasi-position and quasi-momentum representations,
see \cite{ak-gm-rm-prd} and Sec.3.2.

\subsection{Regularisation} 

The aim now is to investigate whether nonzero minimal uncertainties have the
power to regularise the divergencies in $\phi^4$- theory, i.e.
whether the loop summations of perturbation theory, 
such as those in Eq.\ref{tad}, converge on
geometries with minimal uncertainties.
\sn
For the study of the convergence properties of loop summations
the behaviour of the matrix elements of $V$ and $M^{-1}$ for 
large summation indices needs to be established.
\sn
In $\phi^4$- theory on the example   geometries 
${\cal{A}}, {\cal{A}}_1$ or ${\cal{A}}_2$ the
Feynman rule for the Vertex $V$ is specified by applying
the explicit expression Eq.\ref{Crst} for the pointwise 
multiplication to Eq.\ref{Vtuvw} and Eq.\ref{Gamma0}. 
\sn
Recall that in the expression Eq.\ref{Crst} for the $C_{rst}$ only those
terms contribute to the sum for which the 
arguments of all factorials are integers
(Note also that $C_{rst}$ vanishes if $r+s+t$ is 
odd\footnote{In position and momentum space, this is the
integral over the product of three odd Hermite functions.}).
Thus, for fixed $r,s,t$, the number of nonzero terms in the sum 
cannot exceed $r \cdot s \cdot t$
and the $C_{rst}$ can therefore be majorized by:
\be
\vert C_{rst}\vert <
\left(\frac{2}{9\pi L^2}\right)^{\frac{1}{4}}
r s t \mbo  r! s! t! \mbo 3^{r+s+t}
\ee
Using $n!<\sqrt{2\pi n}\mbo  n^n e^{-n} e^{1/12 n}$ 
(from expanding the Gamma function) yields:
\be
\vert C_{rst} \vert < 
\left(\frac{2}{9\pi L^2}\right)^{\frac{1}{4}}
(2 \pi r s t)^{3/2} \mbo (3/e)^{r+s+t} e^{1/12 r +1/12 s +1/12 t} 
\mbo r^r s^s t^t
\ee
Splitting off the non-dominant factors
\be
k(n) := (2\pi n)^{3/2} \mbo (3/e)^n \mbo e^{1/12 n}
\ee
yields in 4 dimensions:
\be
\vert C_{\vec{r},\vec{s},\vec{t}} \vert < 
\frac{2}{9 \pi} \mbo \prod_{i=1}^{4}\mbo L_i^{-1/2} \mbo
k(r_i) k(s_i) k(t_i) \mbo r_i^{r_i} s_i^{s_i} t_i^{t_i}
\ee
The denominator in Eq.\ref{Vtuvw} reflects changes 
arising with the modified geometry. The estimate 
\begin{eqnarray}
[n]_q! & = & \prod_{a=1}^{n} [a]_q = \prod_{a=1}^{n} 
\mbo \sum_{b=0}^{a-1} q^{2b} \nonumber \\
   & = & (1+q^2) (1+q^2+q^4) \cdot ... \cdot (1+...+q^{2(n-1)}) \nonumber \\
   &   & \nonumber   \\
   & > & q^{2(1+2+...+n-1)} = q^{n^2-n} 
\end{eqnarray}
yields for the geometry ${\cal{A}}_2$:
\be
\vert V_{\vec{t}\vec{u}\vec{v}\vec{w}} \vert < 
\frac{4}{81 \pi^2}
\sum_{z_1,...,z_4} \prod_{i=1}^4 L_i^{-1} \frac{
k^2(z_i) z_i^{2z_i} \mbo k(t_i) k(u_i) k(v_i) k(w_i) \mbo
t_i^{t_i} u_i^{u_i} v_i^{v_i} w_i^{w_i} }{
q^{z_i^2-z_i + (t_i^2-t_i+u_i^2-u_i+v_i^2-v_i+w_i^2-w_i)/2}}
\ee
The same majorisation holds in the case ${\cal{A}}_1$ (where there are 
four $q_i$ rather than a $q$), then defining $q:= \mbox{min}(q_1,...,q_4)$.
\sn
The sums are convergent and can be absorbed in a finite 
dimensionless constant $K^4(q)$:
\be 
K(q) := \sum_{z=0}^{\infty} k(z) z^{2z} q^{-z^2 +z}
\ee
to yield for the elementary vertex, using Eq.\ref{Gamma0}:
\be
\vert \Gamma_0(\vec{t},\vec{u},\vec{v},\vec{w}) \vert < 
\frac{\lambda l^4}{4!} \frac{4}{81 \pi^2} 
K^4(q) \mbo \prod_{i=1}^4 \mbo L_i^{-1} \mbo \frac{
k(t_i) k(u_i) k(v_i) k(w_i) \mbo
t_i^{t_i} u_i^{u_i} v_i^{v_i} w_i^{w_i} }{
q^{(t_i^2-t_i+u_i^2-u_i+v_i^2-v_i+w_i^2-w_i)/2}}
\ee
Note that $\Gamma_{\vec{t}\vec{u}\vec{v}\vec{w}}$ is now,
i.e. for $q>1$, highly suppressed for large indices.
It remains to investigate the high index behaviour of the 
Feynman rule $\Delta_0(\vec{a},\vec{b})$ of the free propagator.
\sn
We remark that, e.g. on ${\cal{A}}_1$, the modified inverse propagator
(summed over $i$)
\be
Q^{\prime} :=  \frac{l^2}{\hbar^2}\left(\p_i\p_i +m^2 c^2 
+ \left(\frac{\Delta p_{0i}}{\Delta x_{0i}}\right)^2 \x_i \x_i\right)
\label{irqp}
\ee
is self-adjoint and reads in the Bargmann Fock representation:
\be
Q^{\prime}.\phi(\etab) = 
\left(\sum_{i=1}^4 \left( \frac{(q_i^2+1)^3l^2}{8L_i^2} \etab_i \detab_i + 
\frac{(q_i^2+1)^2 l^2}{8L_i^2}\right) 
+ \frac{m^2c^2l^2}{\hbar^2}\right).\phi(\etab)
\label{qprime}
\ee
Due to 
\be 
\sum_{i=1}^4 \etab_i \detab_i .(\etab_1^{r_1}...\etab_4^{r_4})
= \sum_{i=1}^4 [r_i]_{q_i} \mbox (\etab_1^{r_1}...\etab_4^{r_4})
\ee
it is diagonal, yielding the propagator
\be
\Delta_0^{\prime}(\vec{r},\vec{s}) = 
\left(\sum_{i=1}^4 \left(\frac{(q_i^2+1)^3l^2}{8L_i^2} [r_i]_{q_i} + 
\frac{(q_i^2+1)^2 l^2}{8L_i^2}\right) + 
\frac{m^2c^2l^2}{\hbar^2}\right)^{-1}
\mbo \delta_{\vec{r},\vec{s}} 
\ee
Its nonzero matrix elements rapidly decrease for large indices, due to
the exponential behaviour of $[n]_q=(q^{2n}-1)/(q^2-1)$.
An analogous calculation is possible on ${\cal{A}}_2$.
Recall that the propagator $\Delta_0^{\prime}(\vec{r},\vec{s})$ 
approximates $\Delta_0(\vec{r},\vec{s})$ 
for $\Delta p_{0i} \rightarrow 0$, i.e. for
vanishing minimal uncertainties in momentum and that
it should not differ from $\Delta_0(\vec{r},\vec{s})$ 
in the ultraviolet.
\sn
Nevertheless, an explicit majorisation of the matrix elements of 
the true propagator $\Delta_0(\vec{r},\vec{s})$ is needed. 
Since the Bargmann Fock representation of $Q$, i.e. $M$, is
nondiagonal, the explicit calculation of 
$\Delta_0(\vec{r},\vec{s})$ is rather involved, see 
Eqs.\ref{Qb},\ref{M} and \ref{frfp}.
We can however obtain a majorisation of its crucial high-index behaviour:
\sn
On $F$, which is analytic, the 
operator $Q$ is symmetric and positive definite, 
thus allowing a canonical, lower bound preserving self-adjoint extension.
This so-called Friedrich extension, see e.g.\cite{rs}, 
has a self-adjoint and bounded inverse $Q^{-1}$, defined 
on the entire Hilbert space,
as has every positive definite self-adjoint operator.
\sn
It is crucial that, since $Q^{-1}$ is bounded
$C(q) := \vert \vert Q^{-1}\vert\vert < \infty$, also its matrix
elements $\Delta_0(\vec{r},\vec{s})$ are bounded. 
This follows immediately from the Cauchy Schwarz inequality 
and yields the majorisation:
\be
\vert \Delta_0(\vec{r},\vec{s})\vert \le C(q)
\qquad \forall \vec{r},\vec{s} \in \N^4
\ee
In fact, the lower bound of $Q$ on $F$ is now positive
even in the absence of a mass term,
because, on ${\cal{A}}_1$ and ${\cal{A}}_2$ 
\be
\fsp\left(\phi, \sum_i \p_i\p_i.\phi\right) 
\ge \vert\vert\phi\vert\vert^2
\sum_i (\Delta p_i{}_0)^2 \qquad \forall \phi \in F 
\ee
i.e. technically through what in the language of
quantum mechanics is the existence of minimal uncertainties in momentum
$\Delta p_i{}_0$ (from Eq.\ref{nmun} and \cite{ak-jmp-ucr}), on 
these geometries.
Since the Friedrich extension preserves the lower bound
we obtain self-adjoint and in particular also bounded i.e. 
infrared regular propagators also in the massless case.
More general studies on propagators and  
infrared regularisation are in progress.
\sn
The strong suppression of the matrix elements of the vertex 
for high indices, together with the boundedness of 
$\Delta_0(\vec{r},\vec{s})$ suffices to prove the finiteness of all graphs
in $\phi^4$- theory on the geometries ${\cal{A}}_1$ and ${\cal{A}}_2$:
\sn
Connected graphs $G^{(n)}(\vec{r_1},...,\vec{r_n})$, 
consisting of loop summations over $n_p$ free propagators $\Delta_0$ and
$n_v$ vertices $\Gamma_0$ can be majorised by 
\be
\vert G^{(n)}(\vec{r_1},...,\vec{r_n})\vert < 
C^{n_p}(q) \left(\frac{l}{3} \sqrt{\frac{\sqrt{\lambda}}{\sqrt{6}\pi}} 
\mbo K(q)\sum_{m=0}^{\infty} k(m)\mbo m^m\mbo q^{(-m^2+m)/2}\right)^{4 n_v}
\prod_{i=1}^4 L_i^{-n_v}
\ee
which is convergent due to the summability of the 
sequence $s_n:= m^m q^{-m^2}$, as is readily checked
by the quotient criterion.
\mn
Let us recall that the constants 
$C_{\vec{r},\vec{s},\vec{t}} $ were calculated on the
ordinary geometry ${\cal{A}}$ and have been kept invariant 
while switching on the generalised geometry (i.e. for $q>1$ or the $q_i >1$).
The noncommutative geometry entered into the
Feynman rules through the changes in the 
action of the operators on the fields and their scalar product. 
The modified action of the momentum operators entered into the
calculation of the propagator, yielding in particular an obvious 
infrared regularising effect.
The modified scalar product Eq.\ref{qsp} of Bargmann Fock polynomials entered
into the Vertex, regularising the local interaction. 
\sn
However, to stick to the $C_{\vec{r},\vec{s},\vec{t}}$ of ordinary geometry,
as we did, is only a minimal choice. For generalised Heisenberg algebras 
which imply minimal uncertainties, such as our ${\cal{A}}_1$
and ${\cal{A}}_2$, there is no unitarily equivalent position space 
representation of the commutation relations, which would
uniquely fix the $C_{\vec{r},\vec{s},\vec{t}}$ of the pointwise 
multiplication.

We showed the regularity of the field theory 
without introducing any nonlocality by hand.
But in fact, on noncommutative geometric spaces implying 
minimal uncertainties, the $C_{\vec{r}\vec{s}\vec{t}}$ 
could be modified by hand to some extend, 
introducing an apparent regularising nonlocality, without 
spoiling observational locality.
This is because structure constants which would imply a slight 
nonlocality of the interaction
on ordinary geometry are now to be considered observationally
local as long as the nonlocality introduced is not larger 
than the scale of the minimal uncertainty inherent in the underlying 
geometry, i.e. as long as interaction cannot lead
to an observable nonlocality. This issue needs a further careful 
investigation which will imply the use of maximal localisation states,
see \cite{ak-gm-rm-prd,ak-hh-1}.
We remark that, as is not difficult to check, regularisation on 
${\cal{A}}_1$ and ${\cal{A}}_2$ can be proven along the same lines also
for the infrared modified 
pointwise multiplication $*^{\prime}$ which we mentioned in Sec.3.3.

\section{Summary and Outlook}
In Sec.2 we reviewed and generalised the results of 
Refs.\cite{ak-lmp-bf}-\cite{ak-jmp-ucr} in which is studied
the quantum mechanics on noncommutative geometric spaces
that imply nonvanishing minimal uncertainties in positions and momenta.
Technically, the position and momentum 
operators are symmetric but no longer essentially self-adjoint,
a fact that is crucial in the presence of minimal uncertainties, 
although it is complicating the construction of 
$*$-representations of the Heisenberg algebra.  
Physically, the approach leads to a modified behaviour at very small
and at very large scales, which can be motivated
to arise from gravity and is coinciding with results of string theory.
\sn
In Sec.3 we continued the euclidean field theoretical studies
of \cite{ak-ft,prag}. For two examples of 
noncommutative geometries that imply minimal uncertainties
we worked out the Feynman rules of charged $\phi^4$- theory 
and were now able to prove the finiteness of all graphs. 
The results show, at least in the example of $\phi^4$-theory,
that if gravity or string theory effects induce minimal uncertainties, with
e.g. $\Delta x_0$ of the order of the Planck length, this could
indeed provide a natural regularisation of field theories.
\sn            
Further studies 
in the context of ultraviolet regularisation and microcausality
will use the maximal localisation states to study the locality
properties of generalised pointwise multiplications.
The properties of maximal localisation states \cite{ak-gm-rm-prd,ak-hh-1}
aquire interesting new features in the general $n$-dimensional situation
where the minimal uncertainty gap in the space 
of the $\Delta x_i$ and $\Delta p_j$ can have a complicated structure.
This is being analysed first for the simpler case without minimal 
uncertainties in momenta \cite{ak-gm-rm-2}. 
\sn
We remark that corrections to the commutation relations can imply that e.g.
the $\p_i$ then generate nonlinear transformations of the 
coordinates, which under certain conditions can be
interpreted as the translation of normal coordinate frames 
on a curved space, see \cite{ak-ft,ct}. Further studies on this
`curvature-noncommutativity duality' are in progress.
\sn
For further studies and practical calculations 
on noncommutative geometries other than the two classes
${\cal{A}}_1$ and ${\cal{A}}_2$ which we have covered so far,
it is necessary to construct Hilbert space representations
of the corresponding generalised Heisenberg algebras.
It is not obvious under which conditions the unitary equivalence
of $*$-representations of the Heisenberg commutation relations
(in the sense in which it holds on ordinary geometry) still holds
for generalised Heisenberg algebras. It may not hold for some 
noncommutative geometries in which case the investigation of the
above mentioned dual, curved situation should be interesting. 
One may speculate about a possible relation to horizons or 
nontrivial topology.
\sn
The hope is of course that noncommutative geometric methods 
could provide new techniques for approaching long outstanding 
problems in quantum gravity, as they were outlined e.g. in \cite{isham}. 
On the other hand, as discussed in \cite{ak-jmp-ucr,ak-gm-rm-prd}, 
quantum theory on geometries with minimal uncertainties in positions
could also provide a suitable framework for an effective description 
of nonpointlike particles, which could be strings and, changing scale, 
which could also be compound particles, such
as nucleons in situations in which details of their internal structure
do not contribute, or e.g. various quasiparticles and collective excitations.
Work also in this direction is in progress.

\end{document}